%% file: AA201528048_astroph_lores_wrk1.tex
\documentclass[traditabstract]{aa} 

\usepackage{graphicx}
\usepackage{epsfig}
\usepackage{epsfig}
\usepackage{color}
\usepackage{txfonts}
\usepackage{lscape}
\usepackage{natbib}
\usepackage{longtable}
\bibpunct{(}{)}{;}{a}{}{,} 

\newcommand{\am}[2]{$#1'\,\hspace{-1.7mm}.\hspace{.0mm}#2$}
\newcommand{\HI}{\mbox{H{\footnotesize I}}}

\newcommand{\HIit}{\mbox{H\hspace{0.155 em}{\footnotesize \it I}}}

\newcommand{\MHI}{$M_{\rm HI}$}
\newcommand{\Mz}{$M_{\rm z}$}
\newcommand{\Mstar}{$M_{\star}$}
\newcommand{\MHIMstar}{$M_{\rm HI}$/$M_{\star}$}
\newcommand{\Msun}{$M_\odot$}
\newcommand{\kms}{\mbox{km\,s$^{-1}$}}
\newcommand{\nan}{Nan\c{c}ay}
\def\approxlt{\lower.2em\hbox{$\buildrel < \over \sim$}}
\def\approxgt{\lower.2em\hbox{$\buildrel > \over \sim$}}

\newcommand{\FHI}{\mbox{$F_{\rm HI}$}}

\newcommand{\MHILg}{\mbox{$M_{\rm HI}/L_{\rm g}$}}

\newcommand{\fMHIMstar}{\mbox{$\frac{{M}_{\rm HI}}{{M}_{\star}}$}}

\newcommand{\VHI}{\mbox{$V_{\rm HI}$}}
\newcommand{\Vopt}{\mbox{$V_{\rm opt}$}}

\newcommand{\Wfifty}{\mbox{$W_{\mathrm 50}$}}
\newcommand{\Wtwenty}{\mbox{$W_{\mathrm 20}$}}
\newcommand{\Lr}{\mbox{$L_{\rm r}$}}
\newcommand{\SFR}{{\em SFR}}
\newcommand{\sSFR}{{\em sSFR}}

\newcommand{\SNR}{{\em SNR}}
\newcommand{\SN}{{\em S/N}}

\begin{document}

\offprints{W. van Driel}

\title{NIBLES -- an H{\bf \Large I} census of stellar mass selected SDSS galaxies: \\
I. The \nan\ H{\bf \Large I} survey }
\author{W. van Driel\inst{1,2}
 \and
Z. Butcher\inst{3}
 \and
S. Schneider\inst{3}
 \and
M.D. Lehnert\inst{4}
 \and 
R. Minchin\inst{5}
 \and
S-L. Blyth\inst{6}
 \and 
L. Chemin\inst{1,7,8} 
\and 
N. Hallet\inst{1} 
 \and 
T. Joseph\inst{6} 
 \and
P. Kotze\inst{6,9} 
 \and 
R.C. Kraan-Korteweg\inst{6} 
 \and 
A.O.H. Olofsson\inst{1,2,10} 
 \and 
M. Ramatsoku\inst{6,11,12} 
}
 
\institute{GEPI, Observatoire de Paris, CNRS, Universit\'e Paris Diderot, 5 place Jules Janssen, 92190 Meudon, France 
 \email{wim.vandriel@obspm.fr}
 \and
Station de Radioastronomie de \nan, Observatoire de Paris, CNRS/INSU USR 704, Universit\'e d'Orl\'eans OSUC, route de Souesmes, 18330 \nan, France 
 \and 
University of Massachusetts, Astronomy Program, 536 LGRC, Amherst, MA 01003, U.S.A. 
 \and 
Institut d'Astrophysique de Paris, UMR 7095, CNRS Universit\'e Pierre et Marie Curie, 98 bis boulevard Arago, 75014 Paris, France 
 \and 
Arecibo Observatory, National Astronomy and Ionosphere Center, Arecibo, PR 00612, USA 
 \and 
Astrophysics, Cosmology and Gravity Centre (ACGC), Department of Astronomy, University of Cape Town, Private Bag X3, Rondebosch 7701, South Africa 
 \and 
Universit\'e de Bordeaux, Observatoire Aquitain des Sciences de l'Univers, BP 89, 33271 Floirac Cedex, France 
 \and 
CNRS, Laboratoire d'Astrophysique de Bordeaux-UMR 5804, BP 89, 33271 Floirac Cedex, France 
 \and 
Southern African Large Telescope Foundation, PO Box 9, Observatory 7935, Cape Town, South Africa 
 \and 
Onsala Space Observatory, Dept. of Radio and Space Science, Chalmers University of Technology, 43992 Onsala, Sweden 
 \and 
Kapteyn Astronomical Institute, University of Groningen, Landleven 12, 9747 AV Groningen, The Netherlands 
 \and 
ASTRON, Netherlands Institute for Radio Astronomy, Postbus 2, 7990 AA Dwingeloo, The Netherlands 
}

\date{Received 27/12/2015 ; Accepted  01/07/2016}

\abstract
{To investigate galaxy properties as a function of their total stellar mass, we obtained 21cm \HI\ line observations at the 100-m class \nan\ Radio Telescope of 2839 galaxies from the Sloan Digital Sky Survey (SDSS) in the Local Volume (900$<$$cz$$<$12,000 \kms), dubbed the \nan\ Interstellar Baryons Legacy Extragalactic Survey (NIBLES) sample. They were selected evenly over their entire range of absolute SDSS $z$-band magnitudes (\Mz\ $\sim$ $-$13.5 to $-$24 mag), which were used as a proxy for their stellar masses. Here, a first, global presentation of the observations and basic results is given; their further analysis will be presented in other papers in this series. The galaxies were originally selected based on their properties, as listed in SDSS DR5. Comparing this photometry to their total \HI\ masses, we noted that, for a few percent, the SDSS magnitudes appeared severely misunderestimated, as confirmed by our re-measurements for selected objects. Although using the later DR9 results eliminated this problem in most cases, 384 still required manual photometric source selection. Usable \HI\ spectra were obtained for 2600 of the galaxies, of which 1733 (67\%) were clearly detected and 174 (7\%) marginally. The spectra for 241 other observed galaxies could not be used for further analysis owing to problems with either the \HI\ or the SDSS data. 
We reached the target number of about 150 sources per half-magnitude bin over the \Mz\ range $-$16.5 to $-$23 mag. Down to $-$21 mag the overall detection rate is rather constant at the $\sim$75\% level but it starts to decline steadily towards the 30\% level at $-$23 mag.
Making regression fits by comparing total \HI\ and stellar masses for our sample, including our conservatively estimated \HI\ upper limits for non-detections, we find the relationship log(\MHIMstar) = $-$0.59 log(\Mstar) + 5.05, which lies  significantly below the relationship found in the \MHIMstar\ - \Mstar\ plane when only using \HI\ detections.
}

 
\authorrunning{van Driel et al.}
\titlerunning{NIBLES -- the \nan\ \HI\ survey}
\maketitle

\section{Introduction}\label{intro} 
Understanding the gas cycle in galaxies -- how galaxies acquire, process, and expel their gas -- is the central goal of most studies of galaxy evolution. Our current understanding is that this cycle is a balance between the accretion of gas onto the galaxy, the efficiency of turning the accreted and ``recycled'' gas into stars, and ejecting gas through a coupling of the gas to the luminous and mechanical energy output of stars and active galactic nuclei \citep{bouche10, lilly13}. However, well-studied galaxies, such as our own Milky Way, point to a very different picture \citep{haywood13}. The Milky Way's star formation rate has been roughly constant over the last 9 Gyrs and likely did not drive significant outflows during that period \citep{snaith14, lehnert14}.

This is over a time span during which the cosmological accretion of dark matter was thought to decline by an order-of-magnitude \citep{neistein08,dekel09, dekel13}. To accommodate the high accretion rates onto galaxies relative to their star formation rates, studies often focus on ways of having galaxies drive vigorous massive outflows many times their star formation rates \citep{mitra15}. While this is logical, perhaps it is also important to search for processes that slow down the accretion timescale and the growth of the gas content of galaxies. One plausible way, which is certainly not unique, is to consider the growing angular momentum of accreted gas with decreasing redshift, which has the natural effect of increasing the timescale over which gas is made available for star formation \citep[e.g.,][]{lehnert15}.

The role of \HI\ in galaxy formation and evolution is not yet completely clear. The reservoir of \HI\ gas in galaxies must ultimately feed their star formation \citep{vollmer11}, after cooling and forming molecular clouds. These gaseous disks are very extended, typically beyond the optically bright region of the galaxy \citep{bigiel12}. Since rotation curves are approximately flat out across these outer extended \HI\ disks \citep{vanalbada85}, they dominate the specific angular momentum budget of the galaxy, i.e., the angular momentum per unit mass. This is an important clue to their formation and their longevity. However, to interpret this important clue requires us to have a complete census of the \HI\ content of galaxies. To interpret spatially resolved observations of \HI\ disks, we need to place them into the general context of galaxies. Moreover, although integrated detections of galaxies (at any wavelength) provide only limited constraints on models of galaxy evolution, general demographics of galaxies and gas-phase distributions as a function of mass, environment, and morphological type, are at the moment the only characteristics that models are able to reliably predict. This is simply due to our rudimentary understanding of the physics underpinning galaxy evolution \citep{silk12}.

There are two basic approaches to large \HI\ surveys of galaxies: blind surveys where the sky is scanned to search for detections, and pointed surveys targeting a high number of individual galaxies. Both approaches have their strengths and weaknesses. Blind surveys are best for unbiased detection of \HI-bearing galaxies, even discovering galaxies not previously known \citep{giovanelli13}, and for determining the unbiased comoving density of \HI\ in the local universe \citep[e.g.,][]{zwaan05, martin10}. The disadvantages are that most of the sky is free of \HI\ emission from galaxies, making the surveys time consuming and enabling them to reach only modest depths, and that a sample of \HI-selected galaxies will under-represent populations of galaxies that have low \HI\ content. Moreover, determining detection limits can be tricky and, almost by definition, the upper limits for undetected galaxies lie at similar \HI\ masses as the detections \citep[e.g.,][]{papastergis12}. Thus the upper limits do not add significantly to the analysis of global properties dependent on \HI\ mass, which negates some of the advantages of blind surveys. Pointed surveys have the advantage that the observed sample can be well-selected on particular properties, such as stellar mass or environment, are relatively economical since each pointing guarantees information, whether a detection or an upper limit, and are important for providing multi-variant information. The disadvantages of course are that pointed surveys can be biased in the galaxies they observe, leaving little room for important serendipitous discoveries. 

To aid in the determination of the \HI\ content of galaxies over a wide range of stellar masses, and overcome some of our remaining ignorance of the ``how much'' and ``where is'' of atomic gas, we undertook an \HI\ survey dubbed NIBLES, for \nan\ Interstellar Baryon Legacy Extragalactic Survey \citep{nibles08a, nibles08b, nibles09}.

We observed 2850 galaxies in the local Universe (900$<$$cz$$<$12,000 \kms), selected as uniformly as possible on total stellar mass (for which we used the absolute $z$-band magnitude as a proxy) from the Sloan Digital Sky Survey (SDSS; see, e.g., \citealt{york00}). 
The data were obtained with the 100m-class \nan\ Radio Telescope (NRT; see Sect.~\ref{observations}).

We subsequently supplemented it by four times more sensitive observations of over 150 objects at the 305m Arecibo radio telescope (see Sect.~\ref{observations}). NIBLES, with its uniform selection of galaxies based on total stellar mass, is aimed to complement other recent and/or ongoing large \HI\ surveys in the local volume. These surveys are, in order of the time at which they were started:

\begin{enumerate}
\item {HIPASS:} blind survey at the Parkes 64 m telescope \citep{barnes01}. 
Beam FWHM 14$'$, $rms$ noise level $13~\mathrm{mJy\,beam}^{-1}$ at a velocity resolution of 18 \kms, $-$90$^{\circ}$$<$$\delta$$<$25$^{\circ}$, search range $-$1280 to 12,700 \kms, data taken in 1997-2002 \citep{barnes01}. A total of $\sim$5300 galaxies were detected. The major galaxy catalogs are \citet{meyer04, wong06};

\item {ALFALFA:} blind survey at the Arecibo 305 m telescope. HPBW 4$'$, $rms$ $2.4~\mathrm{mJy\,beam}^{-1}$ at a velocity resolution of 10 \kms, 0$^{\circ}$$<$$\delta$$<$36$^{\circ}$, search range $-$2000 to 18,000 \kms, data taken in 2005-2012, not counting single-horn receiver follow-up observations \citep{giovanelli05a}. A total of $\sim$30,000 galaxies are expected to be detected. The first galaxy catalogs are \citet{giovanelli07, saintonge08, kent08, martin09, stierwalt09}, the subsequent $\alpha$.40 catalog \citep{haynes11} contains 15,855 detections over 40\% of the final survey area; 
the recently uploaded online $\alpha$.70 catalog (http://egg.astro.cornell.edu/alfalfa/data/) contains 25,534 detections over 70\% of the final survey area;

\item {AGES:} blind survey at the Arecibo 305 m telescope of selected small ($\sim$$5^{\circ}\times$$5^{\circ}$) areas sampling different kinds of galaxy environments. HPBW \am{3}{5}, $rms$ $0.6~\mathrm{mJy\,beam}^{-1}$ at a velocity resolution of 10 \kms, search range $-$2000 to 20,000 \kms, data taking started in 2005. A total of 927 objects were detected so far. The galaxy catalogs are \citet{auld06, cortese08, irwin09, minchin10, davies11, taylor12, taylor13, taylor14a, taylor14b, minchin16, keenan16};

\item {GASS:} pointed survey at the Arecibo 305 m telescope \citep{catinella12} of 666 galaxies with stellar masses greater than 10$^{10}$ \Msun\ selected from the SDSS spectroscopic and the Galaxy Evolution Explorer (GALEX) ultraviolet imaging surveys. HPBW \am{3}{3}, mean $rms$ $0.74~\mathrm{mJy\,beam}^{-1}$ at a velocity resolution of 10-21 \kms, 6750$<$\Vopt$<$15,000 \kms, data taken in 2008-2012. A total of 379 galaxies were detected. The final galaxy catalogs are \citet{catinella10, catinella12, catinella13}.
Note: not to be confused with the GASS survey at Parkes of Galactic \HI\ \citep{mcclure09};

\item {EBHIS:} blind survey at the Effelsberg 100 m telescope, of both Galactic and extragalactic sources \citep{winkel10,kerp11,winkel15}. HPBW \am{10}{8}, $-$5$^{\circ}$$<$$\delta$$<$90$^{\circ}$, search range $-$2000 to 18,000 \kms, data taking started in 2009. The current $rms$ for the extragalactic data is $23~\mathrm{mJy\,beam}^{-1}$ at a velocity resolution of 10 \kms\ \citep{floer14}, but observations for a second coverage of the Northern sky are underway which will lower the $rms$ to the level of HIPASS, so a similar sky density of \HI\ detections can be expected.
\end{enumerate}

We omitted the HIJASS blind survey at the 76 m Lovell Telescope at Jodrell Bank which yielded interesting early results, with 424 detections, but was never finished -- see \citet{boyce01, lang03, wolfinger13}, and \citet{davies04} for the three times deeper VIRGO\HI\ survey of the Virgo Cluster.

Furthermore for the 2MASS Tully-Fisher Survey (2MTF) \HI\ data have been published for 1497 targeted galaxies, of which 878 were detected. The galaxy catalogs are \citet{masters08, hong13, masters14b}. These were obtained with the Green Bank Telescope (GBT) and at Parkes with HIPASS; see Sect.~\ref{results} for a comparison between results obtained with these telescopes and the NRT and NIBLES. 

Here, we present the \HI\ survey undertaken at \nan, and limit ourselves to a short discussion of the results. In future papers in this series, we will present the results of our deeper Arecibo \HI\ observations (Butcher et al. 2016a, Paper II, submitted to A\&A), the bivariate luminosity function and \HI\ mass function (Butcher et al., in prep.), stacking of \HI\ spectra of undetected sources (Healy et al., in prep.) and further analyses of the sample. The NIBLES \HI\ data are also used for a comparison with local galaxies with extremely high specific Star Formation Rates (\citealt{lehnert16} and Lehnert et al., in prep.)

In Sect.~\ref{sample} we describe the selection of the observed sample of galaxies, in Sect.~\ref{observations} the observations and data reduction, in Sect.~\ref{results} the results, including a summary of the problems encountered with various SDSS Data Releases (DRs), and in Sect.~\ref{discussion} we present a first, brief discussion.

\section{Sample selection}\label{sample} 
The original sample of about 3000 target galaxies, aimed to be as uniformly distributed over the entire stellar mass range of local galaxies as possible, was selected in 2007 from the SDSS DR5. It should be noted that all SDSS data published here are from the DR9, which was released in 2012, except when explicitly mentioned otherwise (see also Sect.~\ref{results}). Our selection criteria were as follows:

\begin{enumerate}
\item {SDSS data:} must have both SDSS magnitudes and optical spectra in the DR5;

\item {Redshift limits:} must lie within the local volume (recession velocity 900$<$$cz$$<$12,000 \kms) -- we avoid objects nearer than 900 \kms\ to reduce redshift-distance uncertainties and since the automated SDSS photometry has problems with galaxies of very large angular diameter, and we exclude objects farther than 12,000 \kms, because experience has shown this is the effective NRT detection range for all but the most gas-rich, massive galaxies; 

\item {Uniform distribution in absolute magnitude:} uniform sampling of each 0.5 magnitude wide bin in \Mz, with a target of $\sim$150 galaxies per bin; for the least populated bins, at extreme magnitudes, all DR5-cataloged objects were observed;

\item {Observe the nearest objects:} focus primarily on the lowest-redshift objects in each 0.5 magnitude wide \Mz\ bin, as these will have the highest \HI\ flux densities. Similar volumes were sampled for most of the bins (average distance of 30 Mpc up to $-$19 mag, rising to 55 Mpc at $-$21 and 100 Mpc at $-$23 mag); 

\item {No selection on color:} in order to remain all-inclusive in our study of \HI\ properties, we did not want to exclude {\it a priori} objects that could be expected to be gas-poor, such as ellipticals and lenticulars -- their \HI\ properties are not well known as a function of total stellar mass (our selection criterion).
\end{enumerate}

\begin{figure} 
\centering
\includegraphics[width=9cm]{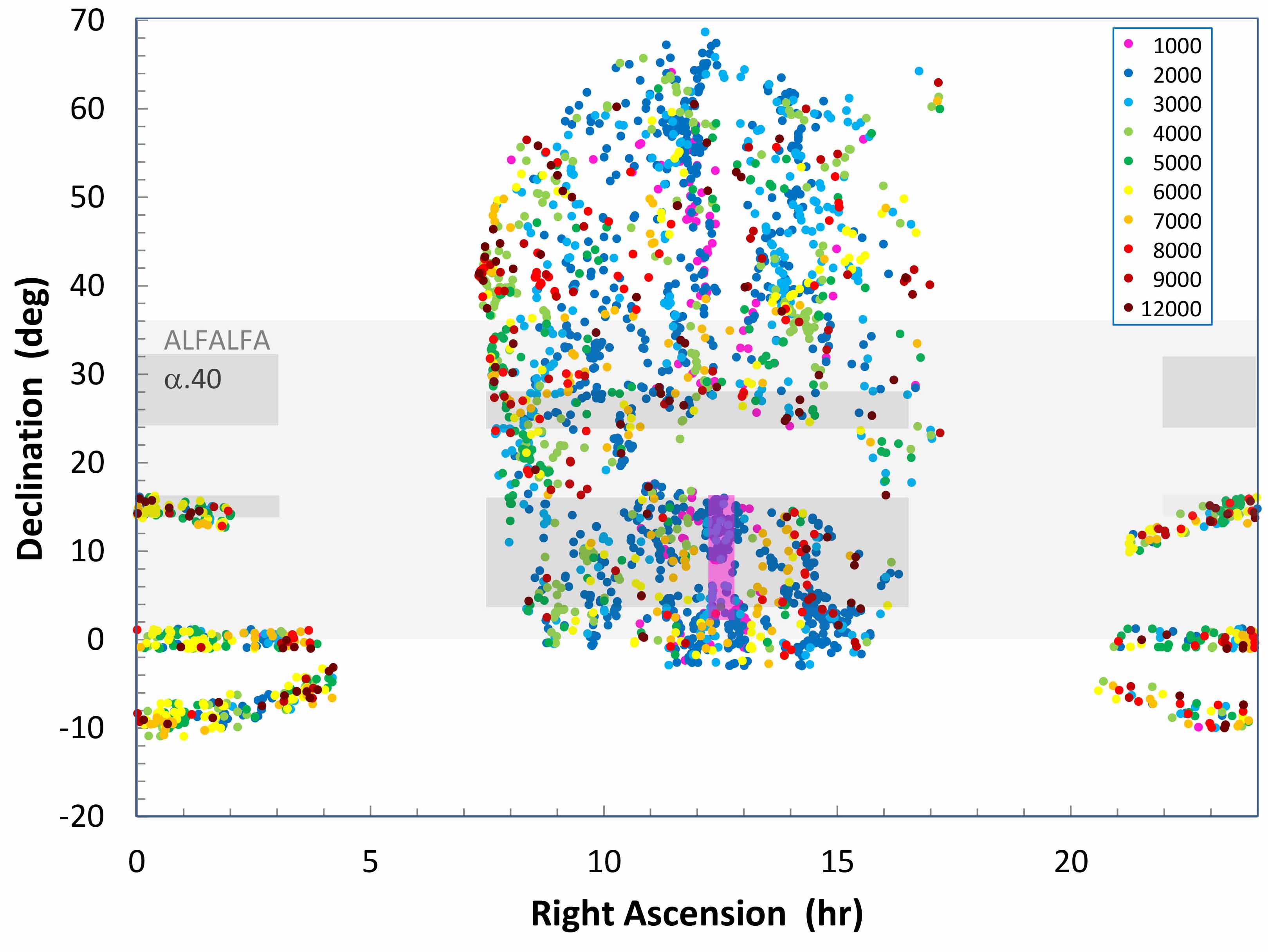}
\caption{Distribution on the sky of the SDSS galaxies observed for NIBLES. The radial velocities of the galaxies have been divided into bins and color-coded accordingly, see the legend. The shaded areas indicate zones that we aimed to avoid, when possible: the pink area indicates the Virgo Cluster volume which only extends out to V = 2000 \kms, and the gray areas indicate the zones to be covered by the ALFALFA blind \HI\ survey: the lighter shade shows the total area, and the darker shade those areas for which results were included in the $\alpha$.40 catalog \citep{haynes11}. }
\label{fig:skydist}
\end{figure}

As far as practicable within the allocated telescope time distribution, when selecting the targets in 2007 we also aimed to avoid (see Fig.~\ref{fig:skydist}) the Virgo Cluster volume, due to the pronounced effects of the cluster environment on the \HI\ properties of its members and the large uncertainties in distances in this region. We also aimed to avoid the declination range to be covered by the blind ALFALFA survey, 0$^{\circ}$ to 38$^{\circ}$ (see the Introduction). The darker gray-shaded area shows the area overlap with the published $\alpha$.40 ALFALFA catalog \citep{haynes11}. 

\section{Observations and data reduction}\label{observations} 

Throughout this paper, all radial velocities given are heliocentric, and all \HI-line related parameters are calculated according to the conventional optical definition (V=c($\lambda$ -- $\lambda_0$)/$\lambda_0$). 
A Hubble constant of H$_{\mathrm 0}$ = 70 \kms\ Mpc$^{-1}$ was used.

The \nan\ Radio Telescope is a 100 meter-class meridian transit, single dish-type instrument located in the center of France (see, e.g., \citealt{monnier03c} and \citealt{vandriel97} for further details). Its collecting area is 6900 m$^2$, equivalent to that of a 94 m diameter parabolic dish. It consists of a fixed spherical mirror (300 m long and 35 m high), a tiltable flat mirror (200 long and 40 m high), and a focal carriage containing two more mirrors and two circular corrugated receiver horns, which can move along a curved rail track. Sources on the celestial equator can be tracked for about 60 min. Due to the E-W elongated shape of the mirrors and the required tilting of the flat mirror for pointing, the N-S beam elongation and telescope gain depend on the observed declination. The HPBW of the elliptical telescope beam  is \am{3}{5} in right ascension, independent of declination, while in declination it is 23$'$ for $\delta$$<$20$^{\circ}$, rising to an estimated 30$'$ at $\delta$= 68$^{\circ}$, the northern limit of the survey \citep[see also][]{matthews00}. The instrument's sensitivity follows the same geometric effect and decreases correspondingly with declination \citep{fouque90}. The minimum system temperature is 35~K at $\delta$=15$^{\circ}$. Flux calibration, i.e., the conversion of observed system temperatures to flux densities in mJy, is determined through regular measurements of a noise diode and periodic monitoring of strong continuum sources by the \nan\ staff; we also made regular observations of \HI\ line calibrator galaxies (see Sect.~\ref{results}). Standard calibration procedures include correction for the declination-dependent gain variation of the telescope (e.g., \citealt{fouque90}). We used an auto-correlator set-up of 4096 channels in a 50~MHz bandpass, with a velocity resolution of 2.6~\kms\ and a velocity coverage of either $-$250 to 10,600 \kms\ or 4750 to 15,600 \kms, depending on source redshift. Data were recorded in the radio convention, in a heliocentric reference frame. The data were taken in position-switching mode, with an elementary integration time of 4+4 seconds per ON-OFF pair. Each observation consists of ``cycles'' of 10/10 pairs of ON/OFF integrations, plus three two-second-long calibrations. For each cycle, the OFF position observation is made along exactly the same portion of the track as the ON position. The observations were made in the period January 2007--December 2010, using a total of about 3450 hours of telescope time. 

We used the standard NRT NAPS software package to identify, flag, and mitigate strong Radio Frequency Interference (RFI), see \citet{monnier03c} for details. The RFI that affects our observations are narrow terrestrial radars in the range 8500-9500 \kms\ and broader, intermittent RFI L3 transmissions around 8300 \kms\ -- for illustrations, see Fig. 3 in \citet{vandriel11}. The RFI-flagging trigger level we use is ten times the $rms$ noise level at full velocity resolution per 40/40 sec ON/OFF cycle of integrations. 
In practice, the radar signals are too strong to mitigate and reliably measure a galaxy \HI\ profile that crosses them. In the case of strong intermittent GPS L3 signals, we exclude the affected integrations from further analysis if they disturb the velocity range of the target.
We used the standard NRT SIR software package to average the two receiver polarizations, perform the declination-dependent conversion from system temperature to flux density in mJy, fit polynomial baselines
(usually third-order, of low amplitude), smooth the data to a velocity resolution of 18~\kms\ and ultimately convert the velocities measured according to the radio convention to the optical system. All \HI\ spectra shown have a heliocentric, optical ($cz$) radial velocity scale.

The \HI\ spectra were reduced using the traditional approach, i.e., by visual inspection of waterfall displays \citep{vandriel11} to verify the quality of the data and the automated RFI flagging with the NAPS package, followed by averaging, baseline fitting, and profile parametrization with the SIR package. Our approach to profile parametrization has been used at the NRT for numerous previous surveys, such as KLUN and KLUN+ \citep[e.g.,][and references therein]{bottinelli92, theureau98, theureau07}. 

We first inspected the entire averaged spectrum, covering a velocity range of about 12,000 \kms, for the presence of what looks like a galaxy \HI\ profile, irrespective of the SDSS redshift. In the rare cases where we found an \HI\ detection at a redshift quite different from the SDSS value we determined its line parameters. We then extracted a velocity range of about $\pm$2000 \kms\ around the SDSS redshift and performed our further analysis of the target's \HI\ line properties within that range. 

To measure the integrated \HI\ line flux a range of velocities was selected that we are confident encompassed the full range of the \HI\ profile. To measure the \Wfifty\ profile width we moved inwards along the profile slopes from the outer edges till the 50\% level of the peak flux density was reached, whereas for the \Wtwenty\ width we measured outwards from the inside \citep[e.g.,][]{lewis83}. The center velocity of the profile, \VHI,  was taken to be the midpoint of the velocity width measured at the 50\% level of the peak flux density.

The NIBLES data reduction was performed before packages for the reliable, completely automated processing of much larger \HI\ data sets were readily available \citep[e.g.,][]{westmeier14, floeer14}. These publications show that the results of classical data reduction are consistent within the quoted uncertainties with the automated results, for profiles with a peak \SNR$>$5. 
We consider our non-automated data reduction procedure adequate for the purpose of NIBLES. Our \HI\ spectra will be made available online in flux density-velocity table format, through CDS, in case other authors wish to carry out their own parametrization procedures. 

Our observing strategy was to first obtain a relatively short observation of each galaxy, using about 40 min of telescope time (resulting in an $rms$ noise level of $\sim$3 mJy at 18 \kms\ velocity resolution), which was repeated in case of weak detections or non-detections, time permitting (see Sect.~\ref{results} and Fig.~\ref{fig:rmsdist}). On average, about 70 minutes of telescope time was used per source.

We obtained higher-sensitivity follow-up \HI\ observations at Arecibo of 90 galaxies not, or only marginally, detected with the NRT and detected 72 of them, with an $rms$ on average about four times lower than the NRT levels (see Sect.~\ref{results}). These results will be described in detail in Paper II.

\begin{figure*} 
\centering
\includegraphics[width=15cm]{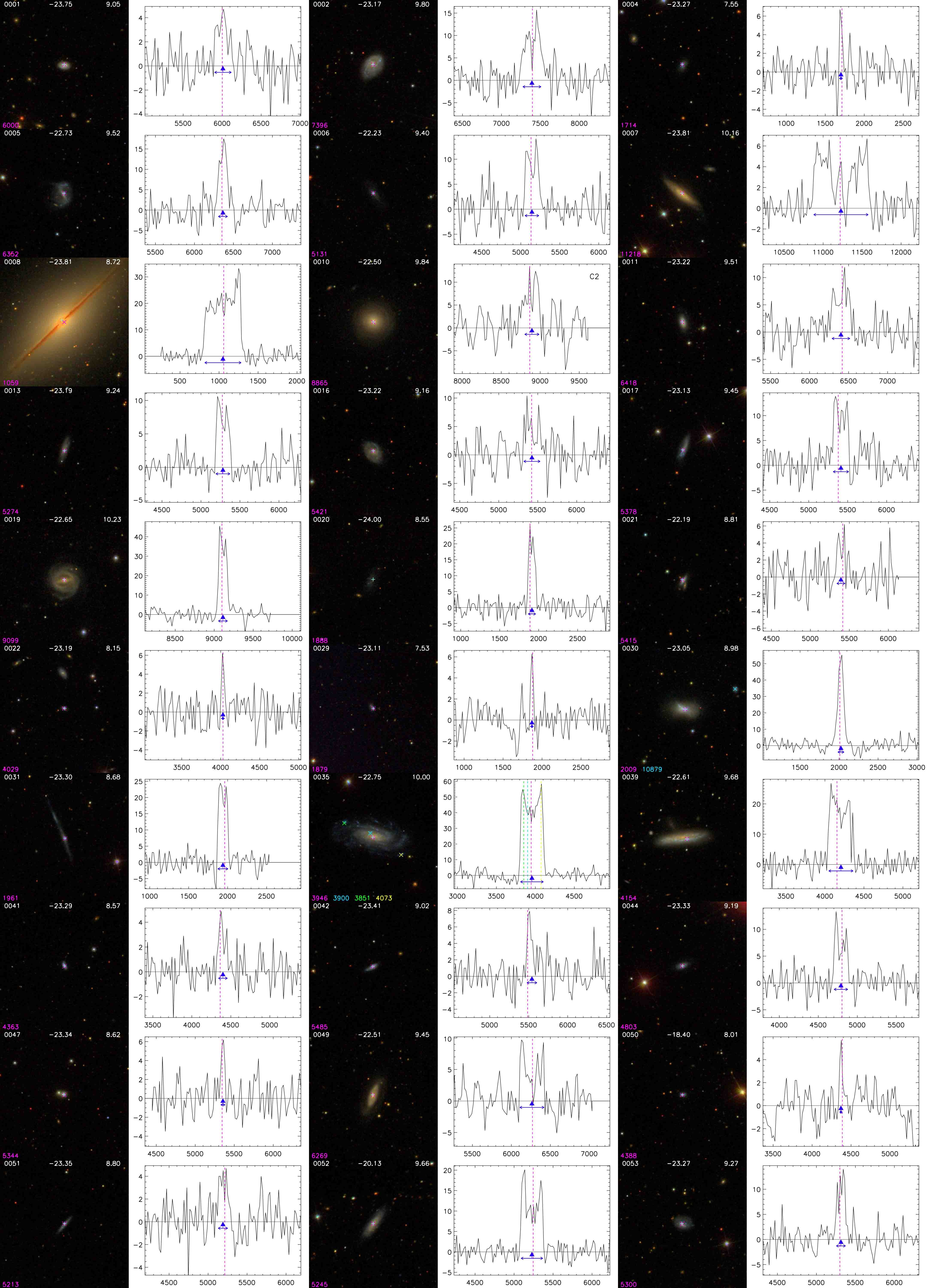}
\caption{{\bf a.} Color images from the SDSS and 21-cm \HI\ line spectra of the galaxies clearly detected at \nan.
The size of each image is \am{3}{5}$\times$\am{3}{5} (i.e., the E-W HPBW of the \nan\ Radio Telescope). Each image is centered on the position of the selected photometric source whose properties are listed in Table~\ref{tab:detections}, the white $+$ sign indicates the pointing center of the \nan\ Radio Telescope, the magenta cross indicates the position of the selected SDSS spectroscopic source whose properties are listed in Table~\ref{tab:detections}, and the crosses of various other colors indicate the positions of the other SDSS spectroscopic positions within the boundaries of the image. Indicated along the top of each image are (from left to right) the NIBLES catalog number of the target galaxy (see Table~\ref{tab:detections}, only available online at CDS), its absolute magnitude in the $z$ band, \Mz, and the logarithm of its total \HI\ mass, log(\MHI) (in \Msun), while indicated in the lower left are the color-coded optical velocities of the SDSS spectroscopic sources in the image. The scale along the horizontal axes of the \HI\ spectra is heliocentric radial velocity ($cz$) in \kms, and the vertical scale is flux density in mJy. Indicated in each spectrum are the central \HI\ velocity (blue triangle) and the \Wfifty\ width of the profile (blue horizonal line), and the SDSS velocity of the selected spectroscopic source (dashed magenta vertical line) and the other sources in the image (dashed vertical lines in colors corresponding to those of the crosses in the image). The velocity resolution is 18 \kms. 
({\it Note: only the first page is shown here
}) 
}
\label{fig:Fig2.1}
\end{figure*}

\begin{figure*} 
\centering
\includegraphics[width=15cm]{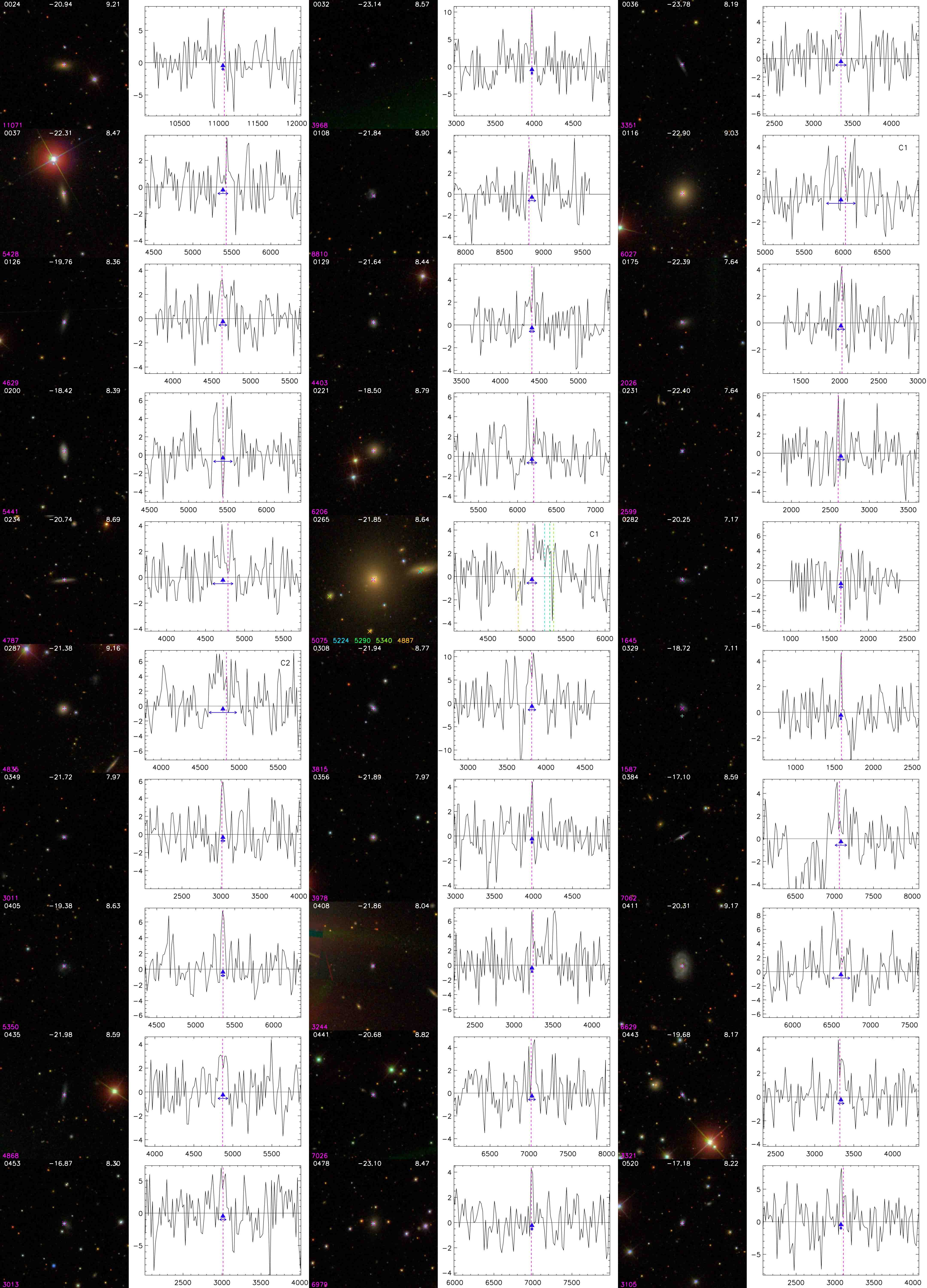}
\caption{{\bf a.} SDSS color images and 21-cm \HI\ line spectra of the galaxies marginally detected at \nan. See Fig.~2a for further details. The properties of the photometric and spectroscopic sources of these galaxies are listed in Table~\ref{tab:marginals} (only available online at CDS).
({\it Note: only the first page is shown here
})
}
\label{fig:Fig3.1}
\end{figure*}

\begin{figure*} 
\centering
\includegraphics[width=15cm]{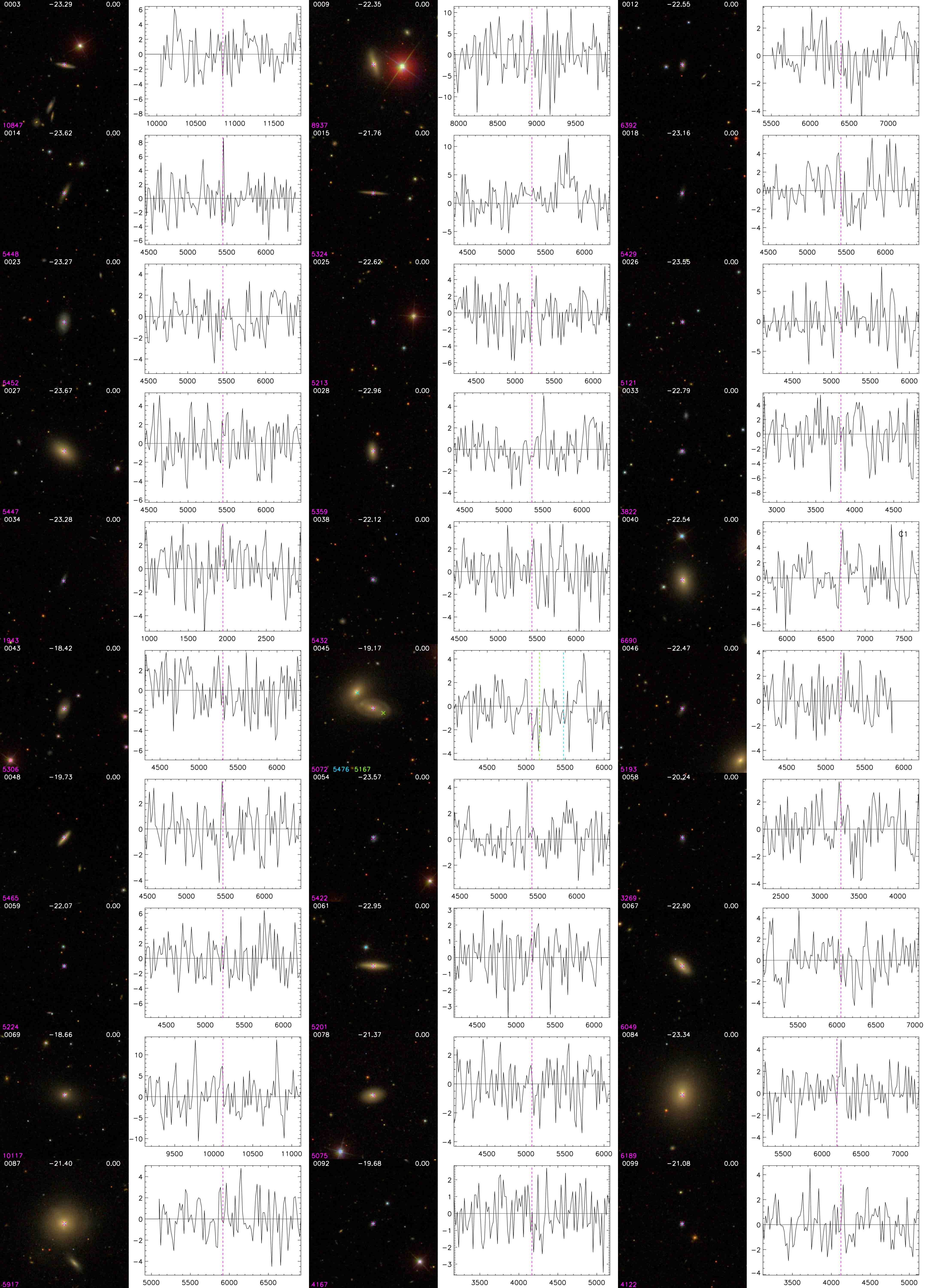}
\caption{{\bf a.} SDSS color images and 21-cm \HI\ line spectra of the galaxies not detected at \nan. See Fig.~2a for further details. The properties of the photometric and spectroscopic sources of these galaxies are listed in Table~\ref{tab:nondetections} (only available online at CDS).
({\it Note: only the first page is shown here
}) 
}
\label{fig:Fig4.1}
\end{figure*} 

On visual inspection, spectra of 83 sources appeared affected by a well-known instrumental baseline ripple \citep[e.g.,][]{wilson09} which can significantly increase the rms noise level. The ripple, related to the presence of a strong continuum source, is caused by radiation reflected between telescope structures which forms a standing wave with a wavelength corresponding to about 115 \kms\ in the case of NRT \HI\ line spectra. As it is a well-defined standing wave, in an FFT deconvolution of a spectrum it is characterized by a narrow peak at always the same position, which can therefore be effectively identified and removed; an inverse FFT then results in a de-rippled line spectrum. We wrote a Python routine to perform this derippling, and will illustrate the derippling process on one of our Arecibo follow-up \HI\ spectra in Paper II. The 12 cases for which the derippling significantly improved the rms noise level and removed the systematic baseline wave pattern have been flagged with a $D$ in Tables~\ref{tab:detections}-\ref{tab:nondetections}.

\section{Results}\label{results} 
Color SDSS images and NRT \HI\ spectra of all our clear \HI\ detections are shown in Fig.~2, marginal detections are shown in Fig.~3 and the non-detections in Fig.~4. 

For the classification of \HI\ spectra as detection, marginal or non-detection, we also considered two signal-to-noise ratios (the peak ratio, \SNR, and the line-width dependent ratio, \SN\ -- see hereafter) but the adjudication was made independently by three of us (ZB, WvD, SES) through visual inspection, after which differently classified sources were discussed in detail, and the final adjudication was made by ZB. Some of the marginal sources would be classified as non-detections in a blind survey, as the median peak signal-to-noise ratio and its standard deviation is 3.0$\pm$0.5 for the sources in this category -- their mean  \SN\ is 3.2$\pm$1.1. 
However, given we know the optical velocities for all our sources, if a peak is coincident with the SDSS velocity, it gives greater credibility to the likelihood of a real signal than if we were searching through velocity space. 

The results from the \HI\ observations and other relevant galaxy properties are listed in Tables~\ref{tab:detections}-\ref{tab:nondetections}. 

Certain physical parameters of the galaxies (MEDIAN total stellar masses and star-formation rates) were taken from the publicly available SDSS ``added-value" MPA/JHU catalogs \citep{brinchmann04, kauffmann03, salim07, tremonti04}, and the remaining optical data are from the SDSS DR9.

\begin{figure} 
\centering
\includegraphics[width=9cm]{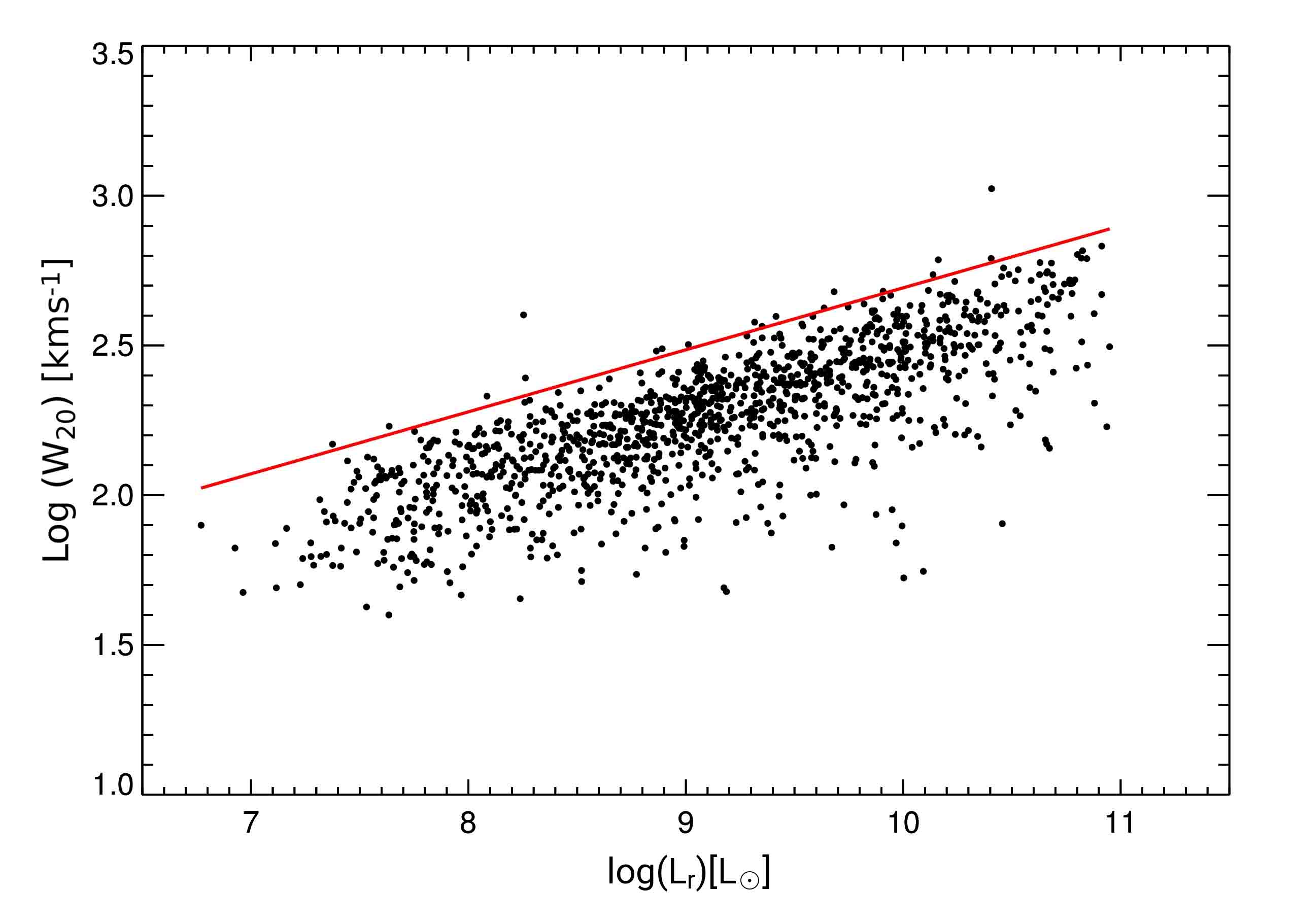}
\caption{NIBLES \Wtwenty\ \HI\ line widths as a function of $r$-band luminosity, \Lr, for all clear, non-confused detections. The red line indicates the upper envelope used in estimating the upper limits to total \HI\ masses of our non-detections (see Table~\ref{tab:nondetections}) as a function of \Lr, log(\Wtwenty) = 0.62 + 0.21 log(\Lr). The two sources (0117 and 2167; i.e., \object{SHOC 30} and \object{NGC 5675}) which lie well above the line have exceptionally broad widths as they are either a merger or a possibly confused detection (see Appendix).}
\label{fig:W20Lr}
\end{figure}

Listed throughout Tables~\ref{tab:detections}-\ref{tab:nondetections} are the following properties of the target galaxies:

\begin{itemize}
\item{Source + flags:} internal NRT target number, which we use for quick object identification throughout the paper. Also indicated are the various flags regarding the SDSS and NRT data (see Sect.~\ref{photoDR9}-\ref{flagged});
\item{RA \& Dec:} Right Ascension and Declination in epoch J2000.0 coordinates, as used for the observations. 
We do not give the latest SDSS catalog name, as its precise coordinates have changed between Data Releases and various databases recognize only older source names;
\item{Name:} common catalog name, other than the SDSS;
\item{\Vopt:} heliocentric radial velocity measured in the optical wavelength domain (in \kms). In case more than one value was listed in the SDSS, the one with the smallest uncertainty and closest match to the \HI\ velocity was chosen;
\item{$g$-$z$:} $g$-$z$ band color, corrected for Galactic extinction, following \citet{schlegel98} (in mag); 
\item{\Mz:} absolute $z$-band magnitude, corrected for Galactic extinction following \citet{schlegel98}; 
\item{\Mstar:} MEDIAN total stellar mass estimates, from the MPA/JHU added-value catalogs (in \Msun);
\item{\sSFR:} specific Star Formation Rate, or \SFR/\Mstar, based on \SFR\ and \Mstar\ from the MPA/JHU added-value catalogs (in yr$^{-1}$);
\item{$rms$:} $rms$ noise level values of the \HI\ spectra (in mJy);
\item{\VHI:} heliocentric radial \HI\ velocity of the center of the \HI\ spectra, taken to be the midpoint of the \Wfifty\ profile width;
\item{\Wfifty, \Wtwenty:} velocity widths measured at 50\% and 20\% of the \HI\ profile peak level, respectively, uncorrected for galaxy inclination (in \kms);
\item{\FHI:} integrated measured \HI\ line flux (in Jy \kms);
\item{\SNR:} peak signal-to-noise ratio, which we define as the peak flux density divided by the $rms$ noise level. For non-detections, the \SNR\ listed is the maximum found in the expected velocity range of the \HI\ profile; 
\item{\SN:} signal-to-noise ratio determined taking into account the line width, following the ALFALFA \HI\ survey formulation from \citet{saintonge07}, 
\SN\ = 1000(\FHI/Wfifty)$\cdot$(\Wfifty/2$\cdot$$R$)$^{0.5}$)/rms, where $R$ is the velocity resolution, 18 km/s;
\item{\MHI:} total \HI\ mass (in \Msun), \MHI\ = 2.36$\cdot$$10^5$$\cdot$$D^2$$\cdot$\FHI, where the galaxy's distance $D$ = $V/H_{\mathrm 0}$ (in Mpc) and $H_{\mathrm 0}$ = 70 \kms\ Mpc$^{-1}$. In the cases of non-detections, 3$\sigma$ upper limits are listed for a flat-topped profile with a width depending on the target's $r$-band luminosity, \Lr, according to the upper envelope in the \Wtwenty-\Lr\  relationship of our clear, non-confused detections (see Fig.~\ref{fig:W20Lr}); 
\item{\MHIMstar:} ratio of the total \HI\ and stellar masses.
\end{itemize}

Our method of determining upper limits to \MHI\ appears to be a fair estimate of the amount of \HI\ flux that could escape detection, as we will show in Paper II using our four times more sensitive Arecibo follow-up observations of \nan\ non-detections.

Estimated uncertainties are given after the values in the tables. Uncertainties in the central \HI\ line velocity, \VHI, and in the integrated \HI\ line flux, \FHI, were determined following \citet{schneider86, schneider90} as, respectively
\begin{equation} \sigma_{v_{HI}} = 1.5(W_{20}-W_{50})S\!NR^{-1}\, (\kms)
\end{equation} 
\begin{equation} \sigma_{F_{HI}} = 2(1.2W_{20}R)^{0.5}rms\, (\kms)
\end{equation} 
where $R$ is the instrumental resolution, 18 \kms, \SNR\ is the peak signal-to-noise ratio of a spectrum and $rms$ is the rms noise level (in Jy). Following Schneider et al., the uncertainty in the \Wfifty\ and \Wtwenty\ line widths is expected to be 2 and 3.1 times the uncertainty in \VHI, respectively.

These formulae give just the expected rms of a signal integrated across a number of channels with uncorrelated Gaussian noise, so this is the minimum expected uncertainty. Fluctuations due to uncertainties in the baseline fit are difficult to quantify, but as the baseline variations are generally rather mild with an amplitude small compared to the peak in the \HI\ signal, we believe the fits are well constrained. In the past, we also tried the more complicated NRT-based formulae of \citet{fouque90} but they yield similar results.

\begin{figure} 
\centering
\includegraphics[width=9cm]{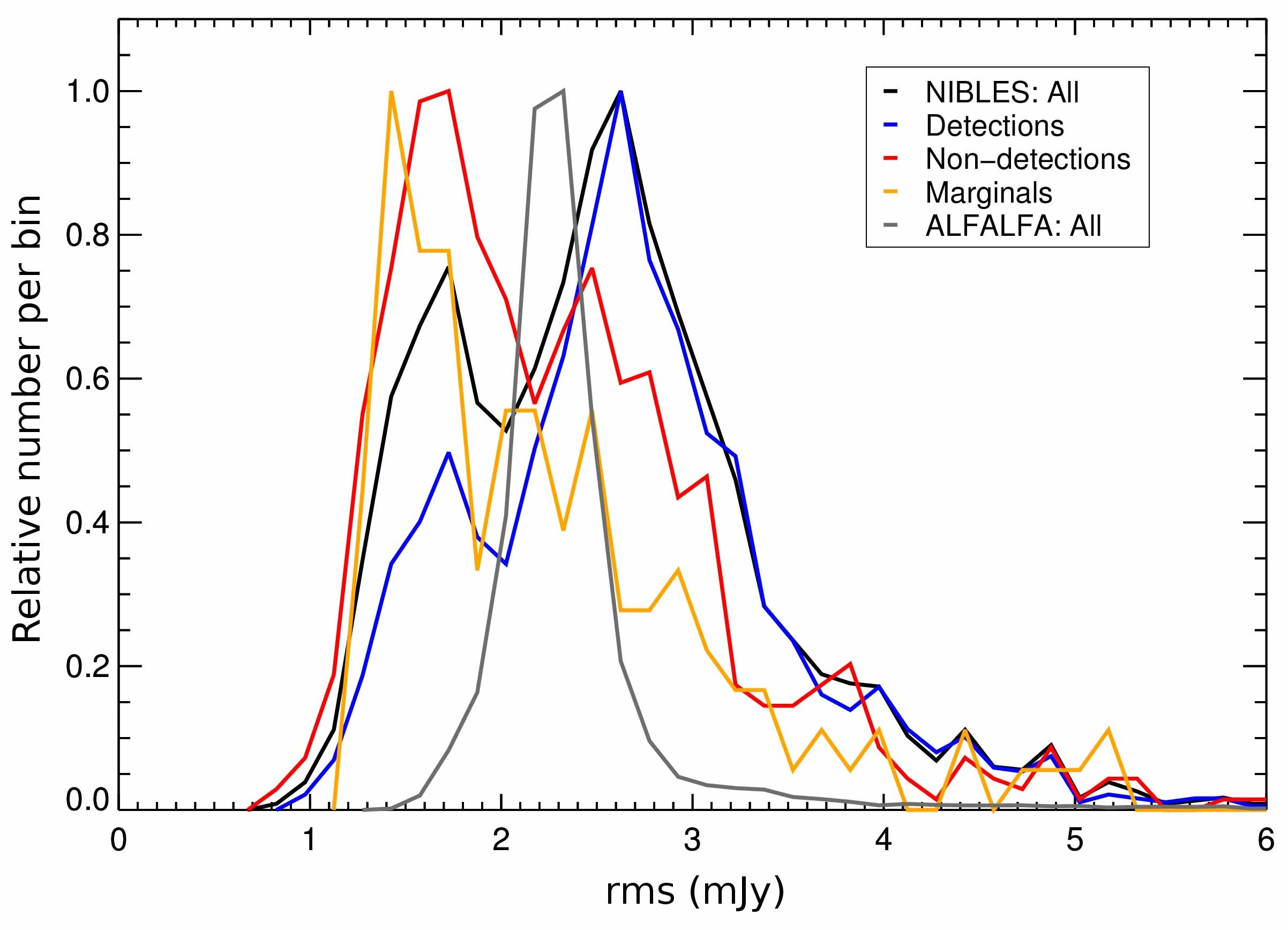}
\caption{Comparison between peak-normalized distributions of $rms$ noise level values below 5 mJy: NIBLES (at 18 \kms\ velocity resolution), all observations (black line, N = 2606), clear detections (blue line, N = 1727), marginal detections (yellow line, N = 176) and non-detections (red line, N = 703), and (dark gray line) all 15,598 ALFALFA $\alpha$.40 catalog detections, at 10 \kms\ velocity resolution \citep{haynes11}}.
\label{fig:rmsdist}
\end{figure}

The distribution of the $rms$ noise levels for our clear detections, marginals, and non-detections is shown in Fig.~\ref{fig:rmsdist} at 18 \kms\ velocity resolution, together with that of the homogeneous, blind ALFALFA survey with its symmetric peak (mean $rms$ 2.3 mJy, at 10 \kms\ resolution). The effect of our flexible observing time strategy is reflected in the peaks around 2.6 mJy (mainly for clear detections after the first run of 40 min) and the lower peaks around $\sim$1.6 mJy of the deeper integrations. The deeper observations permitted us to better constrain the line flux and width for the weak marginal detections and to set lower limits on total \HI\ masses for non-detections.

For a description of the flags used in the tables to point out various issues with the SDSS data, see Sect.~\ref{flagged}. For a description of the 240 cases that were not used for further analysis for various reasons, see Sect.~\ref{excluded}.

\subsection{Photometry update to SDSS DR9}\label{photoDR9} 
Using the SDSS DR5 data upon which the initial galaxy selection was based, as our \HI\ data started accumulating we discovered \citep{joseph08} that a fairly large subset of the detected galaxies showed unprecedentedly high \MHI/$L_{\rm_z}$ ratios, together with very low luminosities. These values for gas content and luminosity did not appear physical, and they were mainly found among the seemingly least-luminous targets: all galaxies with a listed DR5 $g$-band absolute magnitude $M_{\rm g}$$>$-13 mag had \MHILg\ between 10 and 10,000, with a linear increase between $M_{\rm g}$ and log(\MHILg), the apparently less luminous sources being the most gas-rich.

All this made us suspect that the associated magnitudes were in fact severely misunderestimated (a term that captures the complexities involved), which also lead to unrealistically low luminosities of galaxies. Having only the DR5 products at our disposal, it turned out to be infeasible to pinpoint the technical problem(s) involved, to find other measured parameters that would enable the identification of sources with unusable photometry, or how to select the correct magnitudes from the DR5, if those existed. Measuring Petrosian magnitudes ourselves from DR5 FITS images using SourceExtractor \citep{bertin96} we found much brighter magnitudes for suspect sources, which led to normal \MHI/$L_{\rm z}$ ratios \citep{joseph08}.

Since the start of NIBLES, several more SDSS data releases were made public during the observation and data reduction process phase. In particular, data releases 8-12 use an updated photometric processing pipeline (v5\_6\_3) which re-calibrated and re-resolved all the imaging data. In order to take advantage of the new data available, and to address the abovementioned problem with the DR5 SDSS photometry, we updated all the NIBLES photometry to the then current DR9. 

In addition to the updated photometric processing, DR9 has a flux based association between the spectroscopic and photometric information. In the data releases prior to DR8, the only association available was position based, that is, the spectrum was only associated with the nearest photometric source. DR9 incorporates two matches, one based on position and another based on flux. This new matching method provided a much better association between spectroscopic sources and photometric sources with a Petrosian radius large enough to capture most of the flux from a galaxy. However, this did not entirely solve the problem. 

The DR9 photometric information that we decided to use for a galaxy was not always associated with the spectroscopic information via the normally used {\em fluxObjID} in the SDSS database. In two cases, the photometric source associated with the spectroscopic source was instead selected using {\em bestObjID}, because the standard procedure using {\em fluxObjID} returned no result (flag $B$ in the tables).

In another 312 cases, the selection of a photometric source associated with the spectroscopic source using either {\em bestObjID} or {\em fluxObjID} returned no result or one that was obviously missing a significant amount of flux. Here, we queried the photometric database around the NRT \HI\ pointing position (towards the spectroscopic source) and manually selected the photometric source which was centered on the galaxy and had a sufficiently large Petrosian radius to encompass the target galaxy (flag $N$). 

Twenty-three of the observed DR5 sources are not included in the DR9. For these objects, photometric and spectroscopic information from DR7 was used (flag $P$). There are also 20 cases where the DR7 redshift more closely agrees with the \HI\ velocity (flag $Z$). These are frequently low surface brightness (LSB) objects, whose spectra only show weak line features. There are also 18 cases where the DR7 photometry more adequately captured the galaxy's total flux than the DR9 photometry, which have been flagged as $P$ in the tables.

In 25 other cases, both the DR7 and DR9 photometry obviously missed a significant amount of the total flux of the galaxy, or is unreliable due to a bright foreground star (flag $U$).

\subsection{Radial velocity update to SDSS DR9}\label{velDR9} 
Using the SDSS DR9 radial velocities instead of the DR5 values used for the original source selection exposed a number of issues for individual sources, which could not be resolved in all cases; see also the Appendix for details on specific cases.

There are 20 DR9 objects with an unconstrained redshift (error message ``zerr=-1'', nominal V = 1245 \kms). For these we searched the online HyperLeda database and the NASA Extragalactic Database (NED) for independent (i.e., non-SDSS) velocities, and inspected their SDSS spectra and the fits made to the lines. For the five cases with \HI\ velocities we adopted those values instead: three from this paper (sources 1355, 1371, and 1698; i.e., \object{MCG +09-19-160}, \object{CGCG 292-024}, and \object{IC 3612}, respectively) and two from Paper II (sources 0828 and 2140; i.e., \object{ASK 261057} and \object{ASK 082514}). 
For the four \HI\ non-detections with independent optical velocities from the literature we adopted the latter values instead (sources 0998, 1654, 1723, and 2265; i.e., \object{NGC 3156}, \object{PGC 40315}, \object{ASK 77777}, and \object{PGC 3350778}, respectively). We removed the remaining 11 cases (see Table~\ref{tab:excluded}) from our sample as inspection confirmed that no credible velocity can be derived for them. 
 
There are 44 cases where the difference between the DR9 and DR5 velocity exceeds 100 \kms\ (and for 22 of these it is much larger -- at least 300 \kms). For the 16 galaxies with NIBLES \HI-detections we used our \HI\ velocity instead, which is supported by literature \HI\ values for five cases and optical redshifts for four. In all cases these values are consistent with the DR9 velocities, not the DR5 values: 13 sources 0006, 0255, 1217, 1619, 1871, 1894, 2081, 2161, 2333, 2414, 2427, 2562, and 2565;
i.e., \object{PGC 4567836}, \object{PGC 1146688}, \object{NGC 3633}, \object{VCC 423}, \object{UGC 08517}, \object{CGCG 017-048}, \object{PGC 140287}, \object{PGC 4553986}, \object{SHOC 513}, \object{2MASX J21231841+0115175}, \object{ASK 139251}, \object{2MASXi J2340427-092336}, and  \object{2MASX J23411334-1059310}, respectively) from this paper and three sources (0645, 1807, and 2316; i.e., \object{ASK 522849}, \object{PGC 1958740}, and \object{ASK 421256}, respectively) from Paper II. 
For nine cases we used the (average) independent, non-SDSS optical velocity listed in the online HyperLeda database: Sources 0609, 0718, 0824, 1598, 1715, 1717, 1771, 2155, and 2356 
(i.e., \object{KUG 0814+251},  \object{CGCG 263-080},  \object{KUG 0910+433}, \object{VCC 315}, \object{IC 3665}, \object{NGC 4649},  \object{PGC 1132599},  \object{NGC 5644}, and  \object{CGCG 137-019}, respectively). 
For the 19 cases where no independent literature redshifts could be found, we adopted the DR9 velocities in the six cases where the line fits seemed reliable to us (sources 0995, 1712, 2056, 2249, 2397, and 2585; i.e.,  \object{ASK 209206}, \object{ASK 1622}, \object{ASK 400932}, \object{CGCG 221-008}, \object{PGC 3104052}, and \object{ASK 124704}, respectively), whereas the remaining 13 cases (see Table~\ref{tab:excluded}) were removed from further analysis as their DR9 velocities are too uncertain.
 
\subsection{Confused sources}\label{confusion} 
As the HPBW of the telescope used for our \HI\ survey is \am{3}{5} in $\alpha$ and $23'-30'$ in $\delta$, we need to examine a large area on the sky surrounding the target object for the presence of other galaxies whose \HI\ emission might cause confusion with that originating from the intended target.

\begin{figure} 
\centering
\includegraphics[width=9.25cm]{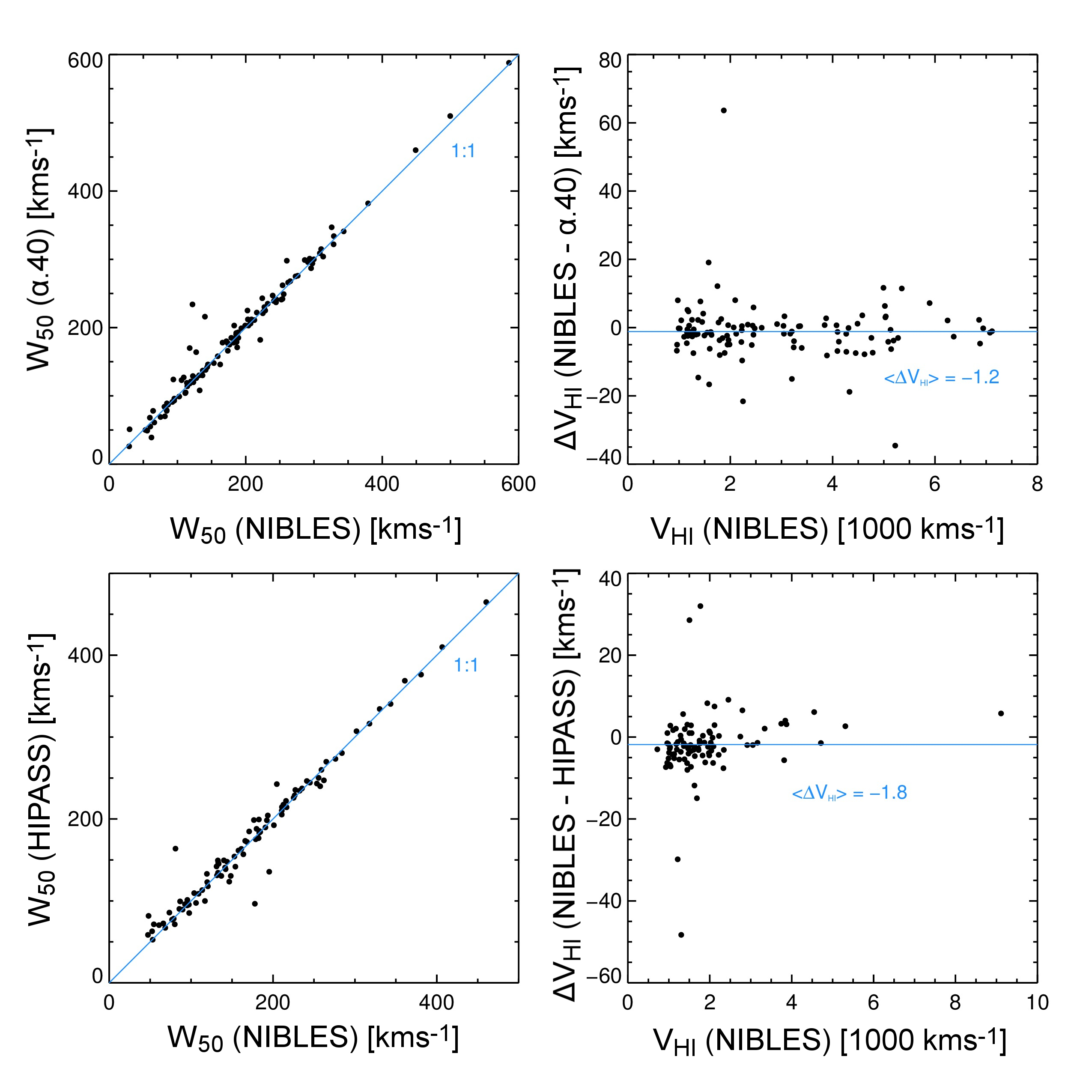}
\caption{Comparison between NIBLES and literature values for central \HI\ line velocities, \VHI, and \Wfifty\ line widths.
Only sources matched in \HI\ velocity and in line width are plotted (see Sect.~\ref{comparison}).
Top panels: comparison with data from the $\alpha$.40 ALFALFA catalog \citep{haynes11}; lower panels: comparison with HIPASS data (\citep{meyer04, wong06}.
Left-hand panels: \Wfifty\ line widths. Literature values as a function of NIBLES widths (in \kms). The solid line marked with ``1:1'' indicates equality between both scales and was only added to guide the eye. Right-hand panels: central \HI\ line velocity. Difference between \VHI\ (in \kms) measured at the NRT for NIBLES and literature values as a function of NIBLES line velocity (in \kms). The horizontal lines indicates the mean velocity differences. 
}
\label{fig:VWcomp}
\end{figure} 

In the pre-SDSS era, this would usually involve a great deal of guesswork, as velocity data for large numbers of galaxies was unavailable. However, since this is a targeted survey of SDSS galaxies, we can make use of the redshift information provided for nearby galaxies to help ascertain whether or not a particular object is a likely source of confusion. As the SDSS does not contain redshift information for every neighboring galaxy, particularly in the case of ellipticals and dwarfs, we also used NED to query for nearby galaxies. 

For each NIBLES source, we first queried NED for all galaxies within a 20$'$ radius and with a recession velocity of less than 12,500 \kms. We then selected galaxies that lay within a box of dimensions 10$'$ in RA and 40$'$ in DEC, centered on the target, and with a systemic velocity located within the \Wtwenty\ width of the \HI\ profile window. 

The latter galaxies constitute our working database of potentially confusing ``secondary'' sources with known redshifts. However, estimating the likeliness of a particular galaxy being a confusion source requires a certain amount of educated guesswork. For example, a small dwarf galaxy near a large targeted blue spiral will contribute only a negligible amount of flux to the integrated observed \HI\ flux, whereas a large gas-rich spiral located just outside the telescope's HPBW is bound to contribute a significant amount of contaminating flux to the observation. 

Next, we extracted an SDSS image of size 14$'$$\times$40$'$ ($\alpha$$\times$$\delta$) centered on each targeted source and superimposed telescope beam contours at levels corresponding to 90\%, 75\%, 50\%, 25\%, and 10\% of the peak sensitivity. We also extracted SDSS images of each secondary source. If a galaxy was contained within the 25\% beam sensitivity contour, it was flagged, regardless of whether or not it was deemed to be a significant contributor of \HI\ flux. Selected galaxies outside this contour were also flagged, but only when they were (at least) comparable in angular size to the target source and considered potential \HI\ contamination sources. We also queried HyperLeda for published \HI\ data on all flagged secondary galaxies. For all potentially confused target galaxies, we estimated the probability of \HI\ profile contamination and flagged them accordingly. We use three flags to indicate the potential contamination level:

\begin{itemize}

\item{C1:} target detection is definitely confused with a secondary galaxy containing comparable or higher \HI\ mass than the target;
\item{C2:} target detection is probably confused, but unable to ascertain how much of the detected flux is likely due to a secondary galaxy;
\item{C3:} target detection may be slightly confused, but the contribution to the detected flux from one or more secondary galaxies is likely insignificant or non-existent. 

\end{itemize}

These three confusion flags are listed in Tables~~\ref{tab:detections}-\ref{tab:nondetections} for all NIBLES sources, whether detected or undetected.

\subsection{Spatially resolved sources}\label{resolved} 
Some of the target galaxies are of sufficiently large apparent size that some \HI\ flux will likely be missed by the NRT beam; these have been flagged as $R$ in the tables. In assessing galaxy sizes and comparing them to the telescope beam, one has to take into account several factors: the telescope is also sensitive to emission outside its HPBW, which is not a ``hard'' beam edge, using isophotal optical diameters such as D$_{\mathrm 25}$ (\citealt{RC3}, HyperLeda) as a selection criterion does not take into account low surface brightness (LSB) disks which usually are relatively \HI-rich, and there is considerable variation in the ratio of \HI\ and optical disk sizes (for various definitions of these two).

\begin{figure}[!h] 
\centering
\includegraphics[width=6cm]{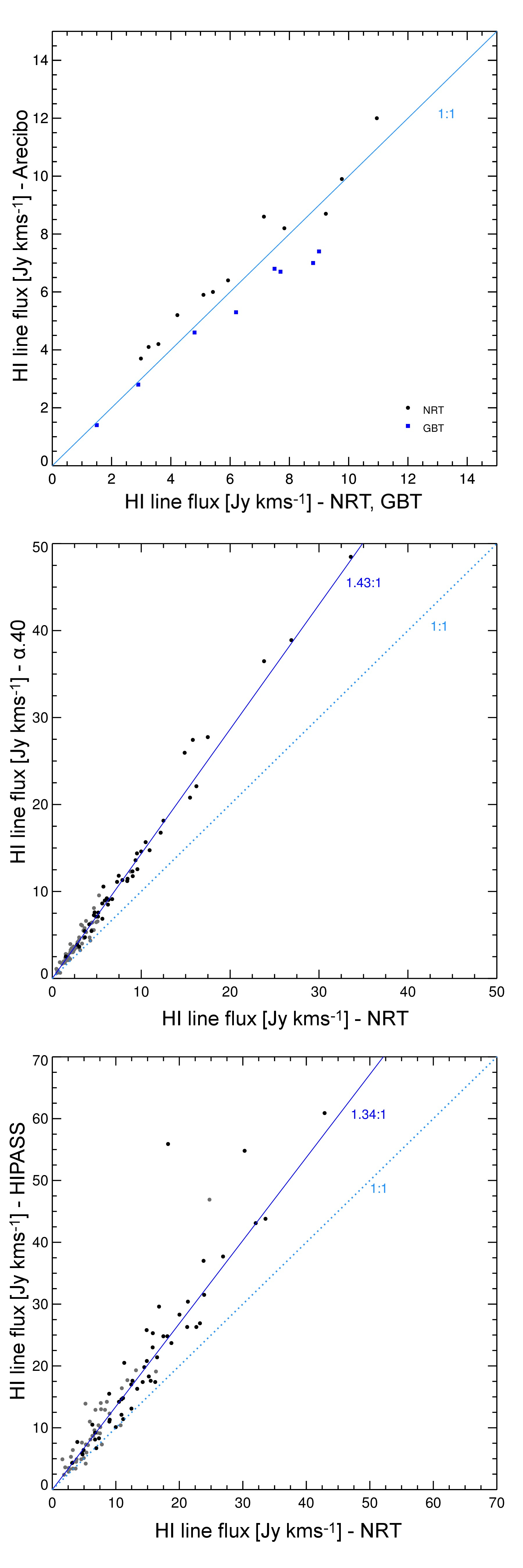}
\caption{Comparison between NIBLES and literature \HI\ line fluxes, in Jy \kms.  
Black dots indicate the sources matched in \HI\ velocity and in line width that were used for the survey flux scale comparisons, whereas the gray dots indicate the other common sources (see also Sect.~\ref{comparison} and Table~\ref{tab:fluxcalibgals}).
In each panel, the solid line indicates the mean relationship between both flux scales, whereas the dotted ``1:1'' line indicates equality between both scales and was only added to guide the eye.
Top pannel: For line flux calibrator galaxies only, as measured with single-feed receivers at the \nan\ Radio Telescope (NRT) for NIBLES and at the Green Bank Telescope (GBT), both compared to Arecibo single-feed receiver fluxes (Arecibo and GBT data are from \citealt{oneil04HI});
middle pannel: Fluxes from the $\alpha$.40 ALFALFA catalog \citep{haynes11} as a function of NIBLES values;
lower pannel: Fluxes from HIPASS catalogs \citep{meyer04, wong06} as a function of NIBLES values.
}
\label{fig:FHIcomp}
\end{figure} 

We therefore adopted the working definition that a galaxy is spatially resolved if it is larger than the NRT HPBW on an SDSS color image. The SDSS images are not very deep, so this should identify objects that have a significant amount of \HI\ flux that will be missed by the NRT. This applies to 85 objects (flag $R$). 

As we intend to keep these in our NIBLES sample for further analysis, we searched the literature for \HI\ observations that can be confidently expected to include their entire \HI\ content (see Butcher et al., in prep.).

\subsection{Cases flagged in the tables}\label{flagged} 
In summary, the following types of cases have been flagged in the tables. 
It should be noted that most of these merely serve to flag technical problems which were ultimately resolved. Only a few indicate sources whose SDSS or \HI\ data have issues which make them unsuitable for use in, e.g., further analysis of relevant global properties of the NIBLES sample, such as \HI\ source confusion ($C1,C2$) and resolution ($R$) or unreliable or underestimated SDSS flux ($U$) -- these flags concern a total of 275 (11\%) out of the 2600 sources in the final NIBLES sample (for the 240 observed sources not included in the final sample, see Sect.~\ref{excluded}).

\begin{itemize}
\item $A1,A2$ (124 and 20 cases, respectively): follow-up \HI\ observations with four times higher sensitivity were obtained by us at Arecibo (see Paper II). Arecibo detections are denoted by $A1$ and non-detections by $A2$.
Here, only their NRT fluxes are listed.; 
\item $B$ (2 cases): the photometric source associated with the spectroscopic source was selected using {\it bestObjID}, because the standard procedure using {\it fluxObjID} returned no result;
\item $C1,C2,C3$ (231,19,185 cases, respectively): \HI\ detection of the target galaxy is confused to a certain degree by another galaxy in the telescope beam. Three different levels are indicated (see Sect.~\ref{confusion}): definitely confused ($C1$), probably confused ($C2$) and only a slight chance of confusion ($C3$);
\item $D$ (12 cases): baseline ripple removed from \HI\ spectrum (see Sect.~\ref{observations});
\item $F$ (175 cases): significant offset between the galaxy center and the SDSS spectral fiber position closest to it. This will usually result in a significant discrepancy between the SDSS velocity and the center velocity derived from the \HI\ profile; 
\item $N$ (465 cases): the selection of a photometric source associated with the spectroscopic source using either {\it bestObjID} or {\it fluxObjID} returned no result or one that was obviously missing a significant amount of flux. Therefore we queried the photometric database around the NRT pointing position and selected the brightest source encompassing the NRT pointing position within its Petrosian radius; 
\item $P$ (25 cases): SDSS DR7 photometry used instead of DR9;
\item $R$ (78 cases): extended source whose integrated \HI\ flux is likely to be underestimated at the NRT (see Sect.~\ref{resolved});
\item $U$ (25 cases): SDSS photometry from DR7 and DR9 almost certainly missed a significant amount of the total flux of the galaxy due to either an obviously too small Petrosian radius or a bright foreground star (see Sect.~\ref{photoDR9});
\item $Z$ (53 cases): SDSS DR7 spectroscopy used instead of DR9.
\end{itemize}

\subsection{Comparison with other \HIit\ surveys}\label{comparison} 
To compare \HI\ line parameters between NIBLES, ALFALFA and HIPASS we matched the coordinates of each NIBLES galaxy to the corresponding galaxy positions in the other surveys using a 30$''$ and 2$'$ radius respectively, i.e., about one sixth of the telescope HPBW.

On comparing our sources to the $\alpha$.40 ALFALFA catalog \citep{haynes11}, we find five ALFALFA sources, observed with a noise level comparable to NIBLES, which were not detected in our data. Of these, three were undetectable at the NRT due to an OFF-beam detection (\object{AGC 180485}, \object{AGC 181493}, and \object{AGC 182461}) and one due to RFI (\object{AGC 320479}). 
The remaining object, source 0023 (\object{KUG 0007+140}), has an ALFALFA mean line flux density at a level (3.6$\sigma$) which should make it faintly detectable in NIBLES, but our four times deeper Arecibo follow-up observation (Paper II) shows a much weaker detection than ALFALFA, at a level undetectable at \nan. 

The mean difference between NIBLES  \HI\ central velocities and data from other large surveys is $-$0.5$\pm$9.8 \kms\ for ALFALFA $\alpha$.40 and -1.8$\pm$8.4 \kms\ for HIPASS, when excluding the \nan\ beam-confused cases (see Fig.~\ref{fig:VWcomp}).

\begin{figure}[!h] 
\centering
\includegraphics[width=9cm]{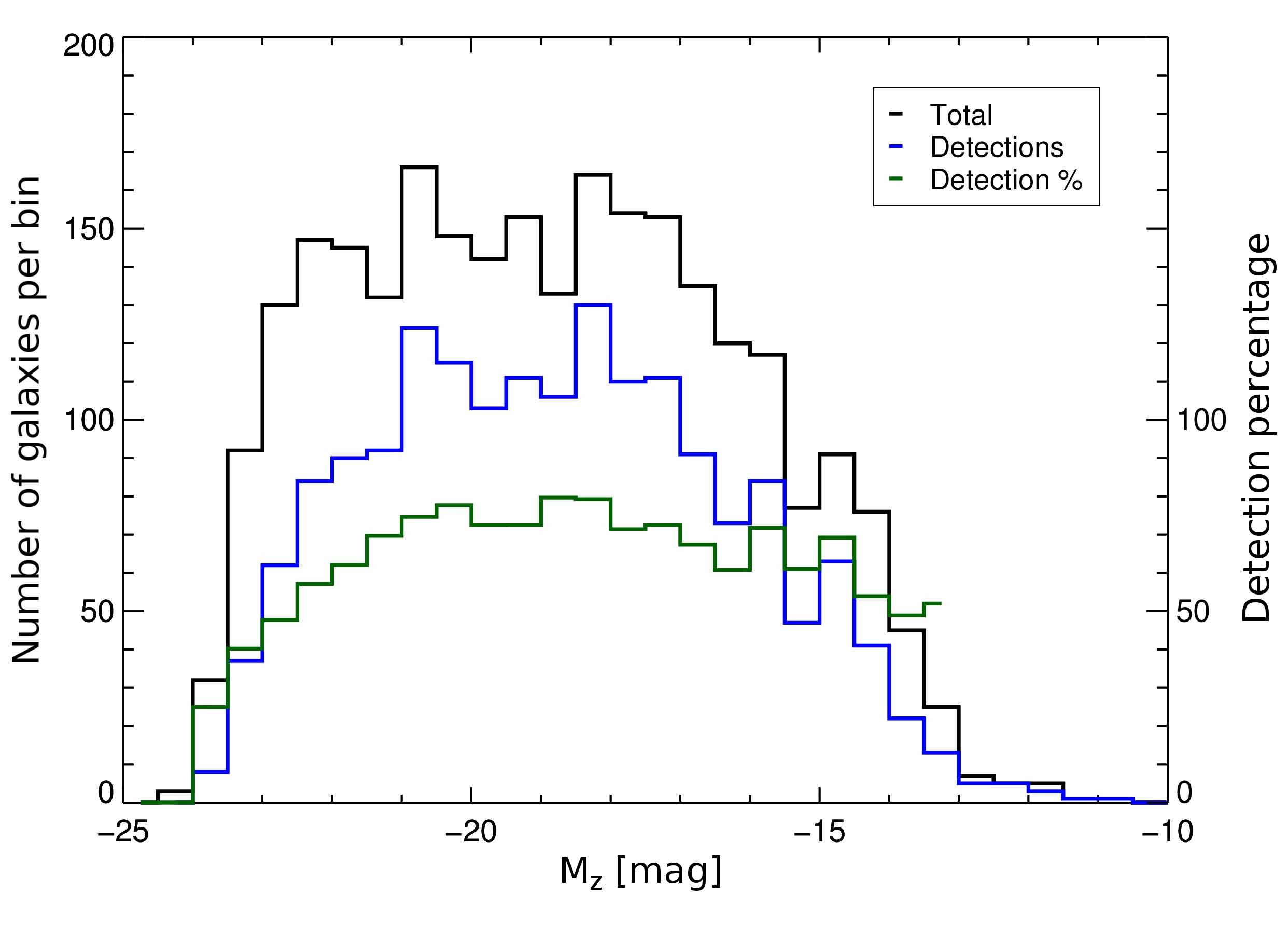}
\caption{Total number of galaxies observed (black line) and detected (blue line) in \HI, either clearly or marginally, per 0.5 wide bin in absolute $z$-band magnitude, \Mz, together with the overall detection percentage per bin (green line). We observed 130-165 objects per half-magnitude bin in \Mz\ over the range of $-23 < M_{\rm z} < -16.5$ mag (the target number was 150), and observed all objects known from the DR5 in the least-filled bins.
}
\label{fig:detstatMr}
\end{figure} 

The six objects with the greatest difference in \Wfifty\ in NIBLES compared to ALFALFA (Fig.~\ref{fig:VWcomp}) are discussed 
in the Appendix (sources 0784, 0793, 0958, 1319, 1970, and 2183, i.e., \object{UGC 4722}, \object{NGC 2743}, \object{PGC 4571034}, \object{PGC 1275866}, \object{NGC 5356}, and \object{LSBC D723-05}, respectively). 
In three cases (0784, 0793, and 1970) the \Wtwenty\ widths are comparable, and the difference in profile shape
only occurs at higher flux density levels. In two cases (0958 and 1319) the NIBLES \Wfifty\ is considerably larger,
whereas in one other (2183) it is much smaller. For the latter three discrepant \Wfifty\ cases no potentially
confusing other source could be identified in the vicinity, nor was any RFI identified in the data.

In order to check the routine NRT flux calibration scale (which is based on continuum sources - see Sect.~\ref{observations}), over the period May 2009 - December 2010 we obtained a total of 64 observations of 12 \HI\ flux calibrator galaxies with very precise New Reference Galaxy Standards for \HI\ Emission Observations from \citet{oneil04HI}, who made observations with two different single-horn instruments -- the L-narrow receiver at Arecibo and the 100-m Green Bank Telescope (see Table~\ref{tab:fluxcalibgals} for the measured line fluxes). These NRT data were obtained and reduced in the same manner as the NIBLES spectra. We found that these line flux scales are consistent (see Table~\ref{tab:fluxcomp} and Fig.~\ref{fig:FHIcomp}): the Arecibo values are on average 1.13$\pm$0.12 (mean and $\sigma$) times the NRT values -- we would like to point out that we did 
{\emph not} use this ratio to rescale our NIBLES line fluxes. The NRT fluxes are also stable over long periods of time: using similar measurements made during the last year we found exactly the same NRT/Arecibo ratio (Kraan-Korteweg et al., in prep.). The other flux ratios are 0.93$\pm$0.24 for GBT/NRT (for two sources only) and 0.99$\pm$0.18 for Arecibo/GBT. The small differences between the flux calibration scales are illustrated in Fig.~\ref{fig:FHIcomp}.

We also cross-checked the fluxes of NIBLES sources with those measured by the HIPASS and ALFALFA surveys (see Introduction for catalog references). 

To reduce noise contamination we only used high peak signal-to-noise ratio NIBLES detections of at least \SNR\ = 12 for the $\alpha$.40 ALFALFA matches and at least 20 for HIPASS. 
We also excluded all NIBLES detections that are definitely or probably confused, or resolved.
To further mitigate effects of beam offsets or confusion sources within the telescope beams in the comparison (we control for confusion in NIBLES, but cannot in the other surveys), we only use galaxies whose difference in central line velocity and \Wfifty\ line width is 20 \kms\ or less between NIBLES and the $\alpha$.40 and HIPASS surveys. This helps ensure we are only measuring differences in flux calibration without contamination by differing velocity offsets. These matching criteria yield 82 flux comparison sources with $\alpha$.40 and 51 with HIPASS (the black dots in Fig.~\ref{fig:FHIcomp}).

Our \HI\ line fluxes were compared using a weighted mean of the ratios of $\alpha$.40/NRT and HIPASS/NRT. Each source was weighted using the inverse of the square of its relative uncertainty in order to give more weight to low uncertainty sources. 

The resulting flux ratios are $\alpha$.40/NRT = 1.45$\pm$0.17 and HIPASS/NRT = 1.34$\pm$0.28, where the uncertainty given is the standard deviation of the weighted mean.
This indicates that the multi-beam survey fluxes are considerably higher than those from 
single-receiver NIBLES.

\begin{figure}[!h] 
\centering
\includegraphics[width=9cm]{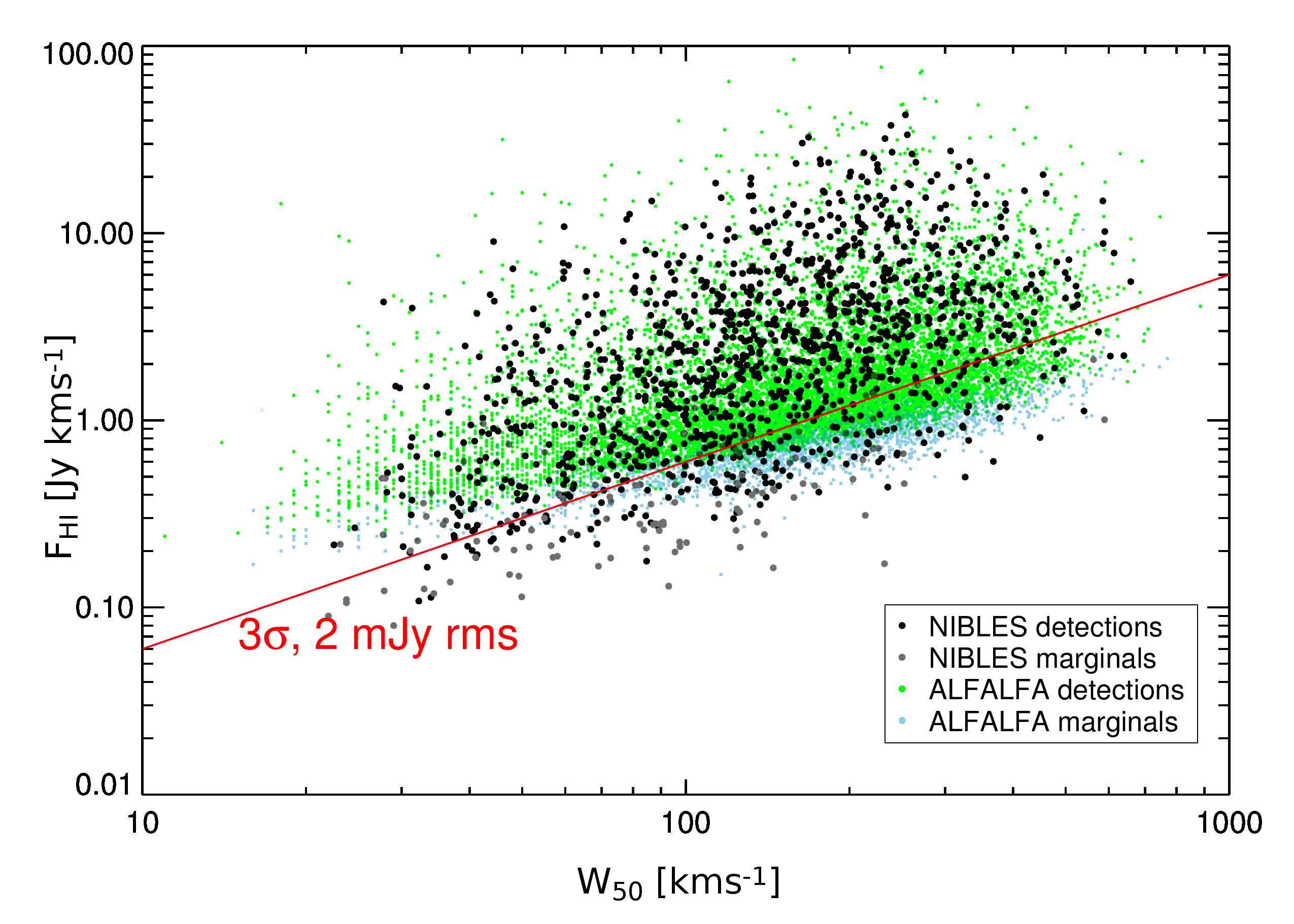} 
\caption{Integrated \HI\ line flux, \FHI\ (in Jy \kms), as a function of \Wfifty\ line width (in \kms). Black dots represent clear NIBLES detections and gray dots marginal detections, whereas green squares are high-quality (their Category 1) $\alpha$.40 catalog ALFALFA detections \citep[from]{haynes11}, and light blue ones their marginal (Category 2) detections. The ALFALFA galaxies shown cover the same velocity range as NIBLES, 900-12,000 \kms. The red line indicates the expected line flux as a function of line width for a flat-topped 3$\sigma$ \HI\ profile and a 2 mJy $rms$ noise level. 
}
\label{fig:FHIW50NIBELSALFA}
\end{figure} 

However, we noted that the $\alpha$.40 catalog ALFALFA flux scale is also higher (though less so than in the case of NIBLES) 
than the Arecibo and GBT \HI\ single-receiver measurements of \citet{oneil04HI} (1.25$\pm$0.16 and 
1.14$\pm$0.09, respectively), and those of the deeper AGES survey (see Introduction) carried out at Arecibo 
with the same multi-beam ALFA instrument (1.20$\pm$0.06). \citet{haynes11} mention that ``some earlier measurements tended to underestimate fluxes for the brightest and more extended sources, a systematic effect for which a correction was applied''; comparing  the results for 2704 sources in the $\alpha$.40 catalog included in earlier ALFALFA 
catalogs (see Introduction), we noted a systematic difference for all sources between the published line fluxes 
which increased with the flux value. This had the effect of aligning the $\alpha$.40 ALFALFA flux scale 
with that of \citet{springob05}, which includes theoretical corrections for \HI\ self absorption and source 
extension, at the expense of agreement with other \HI\ catalogs. 

By fitting the ratio between the $\alpha$.40 line fluxes and those in earlier ALFALFA catalogs, we find a best 
fit of log($F_{\rm HI, ALFALFA \alpha.40}$) = (log($F_{\rm HI, ALFALFA previous}$) $-$ 0.0283)/1.04). 
Using this, we corrected the $\alpha$.40 values back to the older ALFALFA flux scale. We find that the older flux scale is significantly closer to that of NIBLES, with a flux ratio of 1.25$\pm$0.14.

It should be pointed out that the perceived differences in flux scale between multi-beam and
single-receiver observations is not necessarily due to flux calibration differences alone. Compared to measuring total 
line fluxes with single-horn receivers, reconstructing total fluxes from multi-beam receiver data is a rather complex 
process \citep{barnes01} involving choices on regridding, flux conservation, etc. 

We explored this using data from the publicly-available AGES NGC 7332 data cube \citep{minchin10}. For sources chosen to be at a high enough redshift to ensure they are point-like, we measure the sources using two algorithms implemented in the {\sc mbspect} routine in {\sc miriad}. One (which is that used by the AGES survey) fits the position of the source and uses this to reconstruct the flux of a point source at that position using a Gaussian beam model that applies lower weights to pixels further from the position. The second is simply to sum the flux and correct it for the beam size. We carried out this analysis over $5^\prime \times 5^\prime$, $7^\prime \times 7^\prime$, $9^\prime \times 9^\prime$, and $11^\prime \times 11^\prime$ sized regions, centered on the cataloged positions of the sources. We found that the two methods, which should give essentially identical answers for point sources, have markedly different results, with fluxes in the sum-and-correct method increasing with box size well beyond that predicted by a Gaussian beam model. The ratio between the beam-corrected sum over the $9^\prime \times 9^\prime$ region and the beam-weighted reconstruction over the $5^\prime \times 5^\prime$ region was 1.19 on average. 
Although this is exploratory only, it does indicate that changing the way in which total line fluxes are reconstructed from images obtained with single-dish telescopes can significantly change the result.

\begin{figure} 
\centering
\includegraphics[width=9cm]{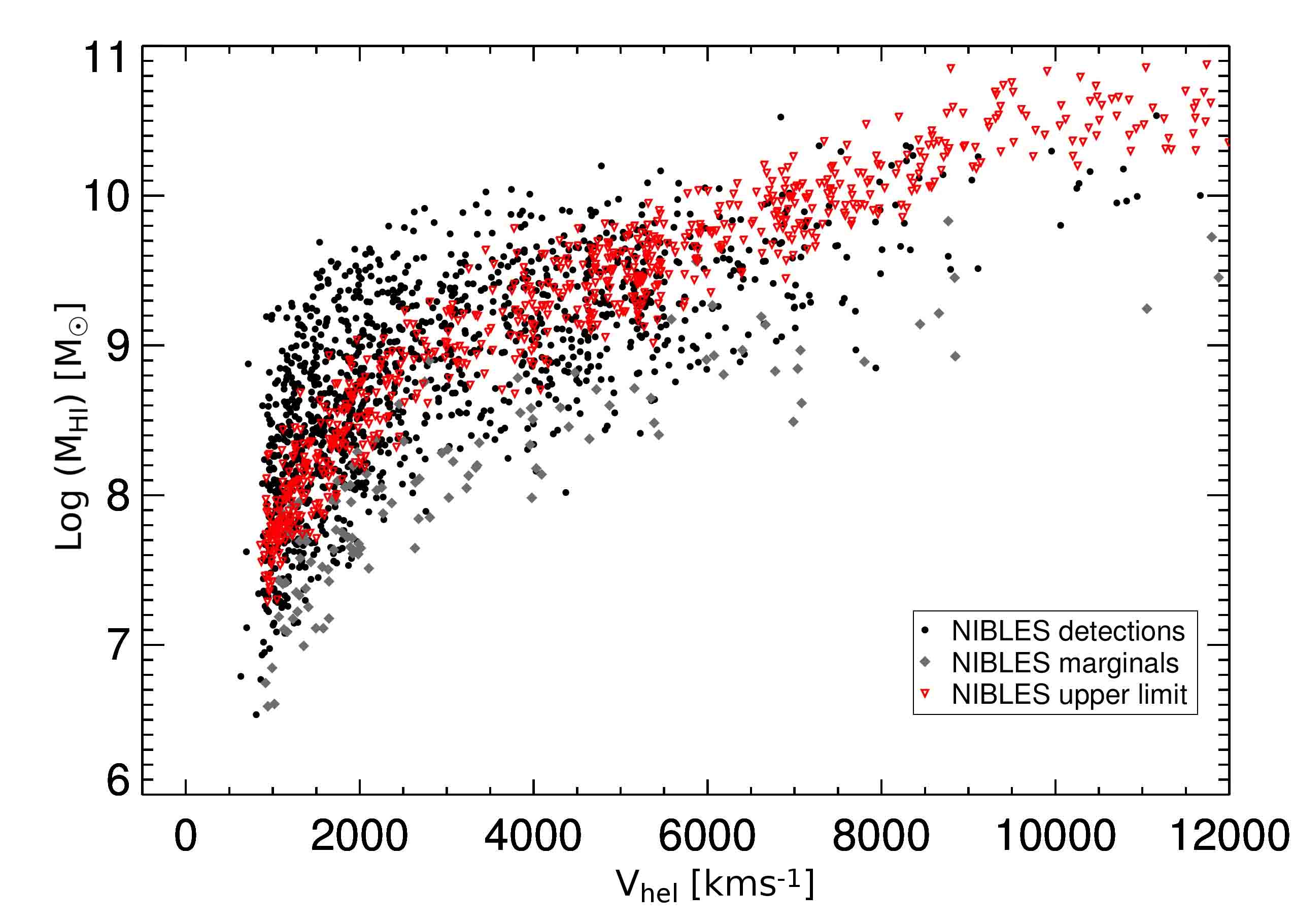} 
\caption{Total \HI\ mass, \MHI\ (in \Msun), as a function of radial velocity (in \kms), i.e., \VHI\ for the detections and SDSS optical values for the non-detections. Black dots represent clear NIBLES detections, gray diamonds marginal detections, and open red triangles estimated upper limit for non-detections. Excluded were all NIBLES detections which are either resolved (flag $R$ in the tables), or definitely or probably confused by another galaxy within the telescope beam (flags $C1$ and $C2$). The relatively high estimated upper limits for non-detections are quite conservative, as they are based on the largest \Wtwenty\ line widths measured as a function of $r$-band luminosity (see Sect.~\ref{results}). 
}
\label{fig:MHIVNIBLES}
\end{figure}

\begin{figure} 
\centering
\includegraphics[width=9cm]{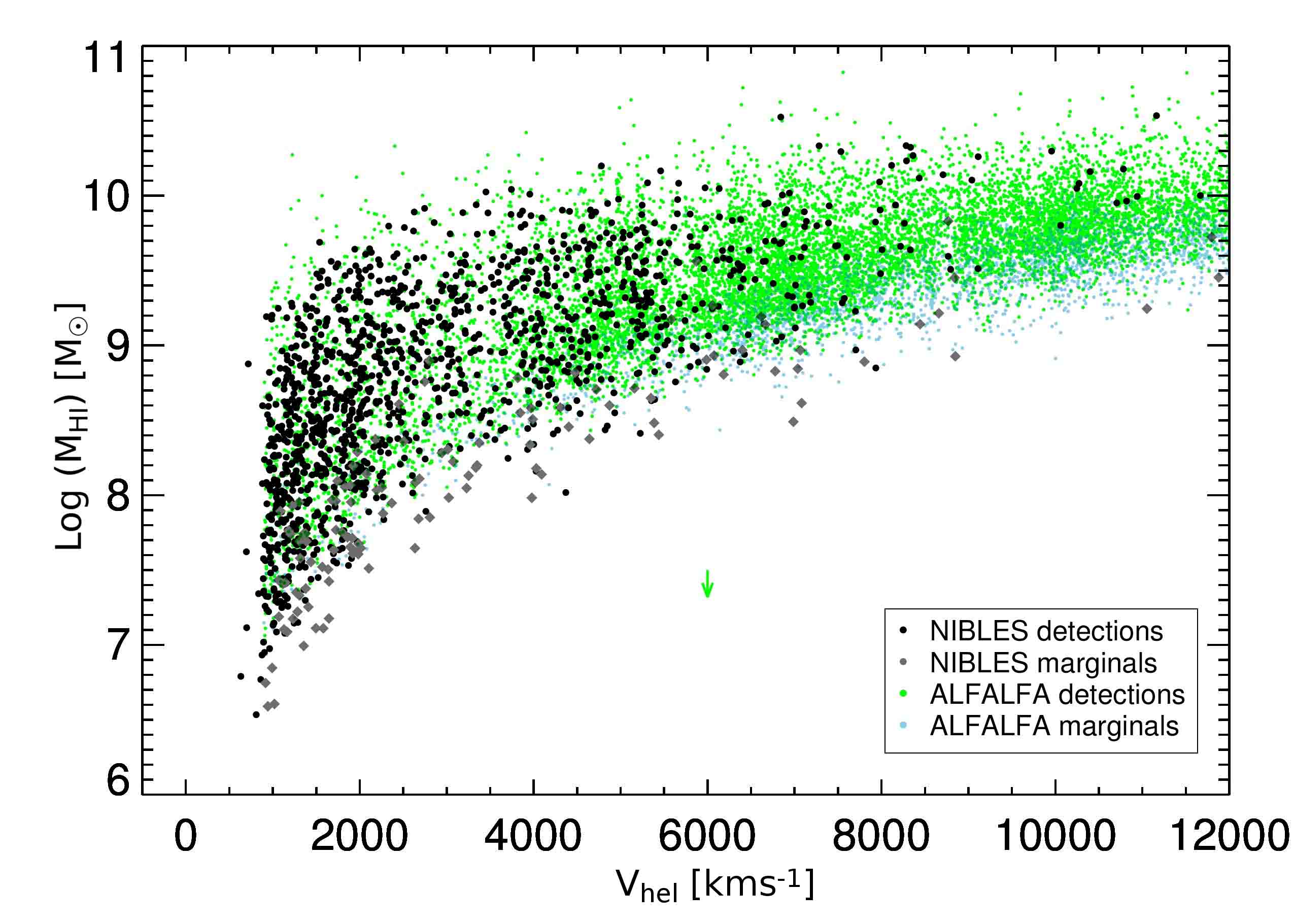} 
\caption{Total \HI\ mass, \MHI\ (in \Msun), as a function of radial velocity (in \kms) -- as in Fig.~\ref{fig:MHIVNIBLES}. Shown besides the NIBLES detections and marginals are the ALFALFA data from \citet{haynes11}, where green squares represent high-quality (Category 1) detections, and blue ones the weaker (Category 2) ones. The \HI\ masses of the ALFALFA detections were calculated in the same way as for the NIBLES sources, using simply a distance of $D$ = $V/H_0$, where the adopted Hubble constant is $H_0$ = 70 \kms\ Mpc$^{-1}$. For clarity, the velocity range plotted is limited to the NIBLES search range limit of 12,000 \kms, whereas the ALFALFA search range continues till 18,000 \kms. The green vertical arrow left of the legend of 0.16~dex in log(\MHI) indicates the difference between the on average 1.45 higher ALFALFA $\alpha$.40 catalog fluxes and the NIBLES values (see Sect.~\ref{comparison}).
}
\label{fig:MHIVNIBLESALFA}
\end{figure} 


\subsection{CRUMBS: stacking of NIBLES non-detections}\label{CRUMBS} 
The NIBLES sub-project named CRUMBS (Characterizing Radio-Undetected Masses in Baryonic Surveys), includes an investigation of all NIBLES non-detections using the spectrum stacking technique. Preliminary results are presented in \citet{blyth09} and further results will be published as part of this series (Healy et al., in prep.). 

\subsection{Sources not included in the final NIBLES sample}\label{excluded} 
Data were obtained for a total of 240 sources among the 2840 observed that are not part of the final NIBLES sample of 2600 objects. They are not listed in the tables as their data will not be used for further analysis of the results of our surveys. They were excluded for the following reasons, listed in order of the frequency of occurrence:

\begin{itemize}
\item{OFF-beam detection (95 cases):} the \HI\ line profile of a galaxy detected in the OFF-beam, which appears as a negative signal in the reduced spectrum, lies (partly) in the velocity range where \HI\ emission from the target is expected to occur, or is in fact observed;
\item{High velocity, out of \HI\ search range (28 cases):} these sources have radial velocities in DR9 which are significantly higher than their DR5 values, and which consequently lie outside our velocity search range; curiously, 17 of these have reported redshifts in the narrow range 23,100 to 23,700 \kms; 
\item{Telescope problem (20 cases):} telescope malfunction which makes the data unusable (mainly pointing and receiver problems);
\item{Not a galaxy (18 cases):} originally classified as galaxies in DR5, but in DR9 17 were reclassified as stars, and one appears to be a cosmic ray;
\item{Unreliable redshift (14 cases):} conflicting redshift information between DR7 and DR9. Only some have warnings of unreliable line fits. These sources were also all undetected in \HI\ so there is no independent measure of their redshift (see Table~\ref{tab:excluded} and Sect.~\ref{velDR9});
\item{Unconstrained redshift (12 cases):} these sources have a listed uncertainty of z$\pm$1 in their SDSS redshift, and they also all have the same reported radial velocity of 1245 \kms. They are listed in Table~\ref{tab:excluded} (see also Sect.~\ref{velDR9});
\item{RFI (10 cases):} Radio Frequency Interference (RFI) is present in the velocity range around the SDSS velocity; 
\item{Baselines bad (9 cases):} unstable baselines which made it impossible to verify if an \HI\ line signal was present; 
\item{Low velocity (6 cases):} radial velocity in SDSS DR9 $<$850 \kms, well below the 900 \kms\ lower velocity limit of our sample (see Table~\ref{tab:lowvels}).
\item{Incorrect velocity search range selected (3 cases):} these sources had an observed \HI\ velocity search range between 0 and 10,000 \kms\ whereas their current SDSS redshifts are over 10,000 \kms.
\end{itemize}

\section{Discussion}\label{discussion} 
Further discussion and analysis of the data presented here will be given in future papers in this series (e.g., Paper II; Butcher et al., in prep.; Healy et al., in prep.). 

Our goal was to observe a total of 3000 galaxies in the local Universe, distributed as uniformly as possible over the full range in absolute $z$-band magnitude detected in the SDSS, without selection on color. The final \Mz\ distribution is shown in Fig.~\ref{fig:detstatMr}, which contains all sources observed and detected, both clearly and marginally. For some galaxies the redshift and/or photometry changed as a result of moving from the SDSS DR5 data used to select them to the DR9 data used for further analysis (see Sect.~\ref{results}). 

\begin{figure} 
\centering
\includegraphics[width=9cm]{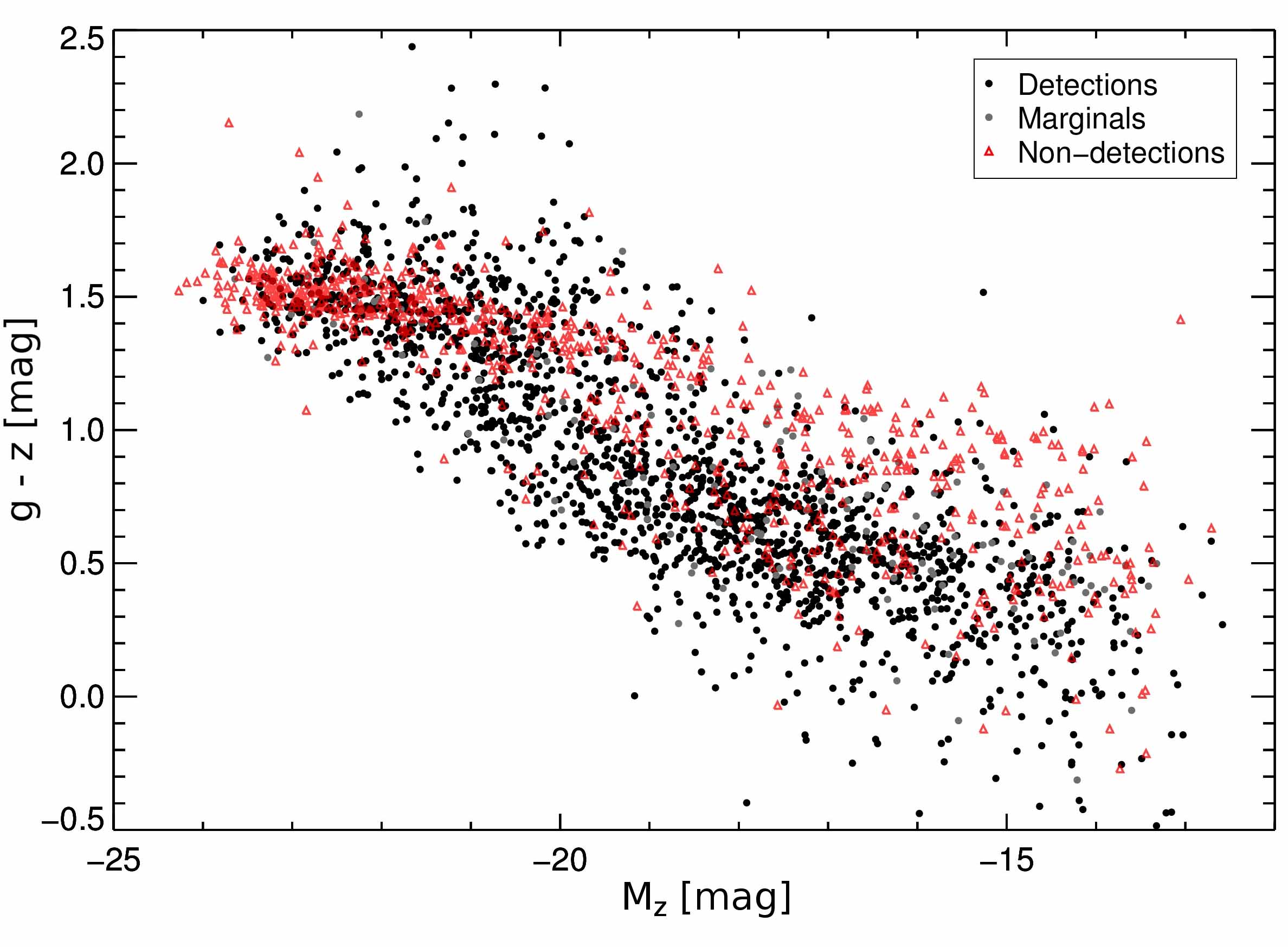}
\caption{Integrated $g$-$z$ color, in mag, as a function of absolute $z$-band magnitude, \Mz. All data were corrected for Galactic extinction following \citet{schlegel98}. Black dots represent clear NIBLES detections, gray dots marginal detections, and red triangles estimated non-detections. 
}
\label{fig:g-zMzNIBLESALFA}
\end{figure} 

We observed 130-165 sources per half-magnitude bin over the absolute magnitude range $-23 < M_{\rm z} < -16.5$ mag (the target number was 150), and observed all objects known from the DR5 in the least-filled bins (see Fig.~\ref{fig:detstatMr}). For \Mz$>-21$ mag the overall detection rate is high, at about the 75\% level, and rather constant, but it starts to decline steadily for more luminous galaxies, towards the 30\% level at $-$23 mag (see also Fig.~\ref{fig:g-zMzNIBLESALFA}).

Shown in Fig.~\ref{fig:FHIW50NIBELSALFA} is the integrated \HI\ line flux, \FHI, as a function of \Wfifty\ line width for the 1733 clear detections and 137 marginal detections, together with the line indicating the flux expected for a flat-topped 3$\sigma$ detection with a 2 mJy $rms$ noise level. For comparison, the same data are shown for the 11,941 ALFALFA high-quality detections (their Category 1) and 3100 marginal detections (their Category 2) from \citet{haynes11}.

As noted in Sect.~\ref{observations}, the $rms$ noise level is not constant for the entire NIBLES sample (see Fig.~\ref{fig:rmsdist}), with more observing time spent on weak marginal detections and non-detections, telescope time permitting; the mean $rms$ for the sample is 2.1 mJy.
The comparable overall $rms$ noise level of the NIBLES and ALFALFA surveys is reflected in their similar \FHI-\Wfifty\ distribution (Fig.~\ref{fig:FHIW50NIBELSALFA}), where also the effect of longer integrations for weak NIBLES marginal detections can be noted. The higher velocity resolution of the ALFALFA data, 10 \kms\ compared to 18 \kms\ for NIBLES, has resulted in a number of very narrow ALFALFA detections, more extreme than found by us.

The distribution of the total \HI\ masses as a function of radial velocity is shown in
Figs.~\ref{fig:MHIVNIBLES} and \ref{fig:MHIVNIBLESALFA}. The former shows both the clear and the marginal NIBLES detections as well as the estimated upper limits of the non-detections, whereas the latter shows the clear and the marginal detections of both NIBLES and ALFALFA. Excluded in both figures were the NIBLES sources which are definitely or probably confused by another galaxy within the telescope beam (flags $C1$ and $C2$ in the tables). For the sake of clarity, the velocity range is only plotted to the NIBLES limit of 12,000 \kms, whereas ALFALFA detections continue out to 18,000 \kms. 

Figure~\ref{fig:MHIVNIBLES} shows that the estimated upper limits to the \HI\ masses of NIBLES non-detections are quite conservative, as they are based on the upper envelope of the \Wtwenty\ line widths measured as a function of $r$-band luminosity (see Sect.~\ref{results}). At the highest velocities (V$>$8000 \kms), where in general the most luminous sources are located with the greatest expected line widths, the upper limits even tend to be higher than the NIBLES detections at the same redshift.

As can be seen in ~\ref{fig:MHIVNIBLESALFA}, the NIBLES SDSS sources were selected at the lowest velocities in their \Mz\ bins, as far as practicable, whereas ALFALFA is a blind \HI\ survey, bound to detect high numbers of relatively high \HI\ mass objects at larger distances than the SDSS sources selected for NIBLES. The ALFALFA line fluxes from the $\alpha$.40 catalog are on average 1.45 times higher than our \nan\ values due to a flux calibration difference (see Sect.~\ref{comparison}), which corresponds to a difference of 0.16~dex in log(\MHI) -- see the vertical arrow in the plot.

The integrated $g$-$z$ color as a function of absolute $z$-band magnitude is shown in ~\ref{fig:g-zMzNIBLESALFA} with different symbols for NIBLES clear detections, marginal detections, and non-detections.
NIBLES sources were selected on \Mz, irrespective of color. The distribution shows the well-known ``red sequence'' and ``blue cloud'' loci, with a mixture of detections and non-detections in both. The highest concentration of non-detections occurs among the most luminous red systems, but these are not all elliptical systems and \HI\ detections do occur. Among the low-luminosity systems (at log(\Lr) 7.8-8.5) the non-detected dwarfs are predominantly red. We will study the underlying \HI\ properties of undetected galaxies using four times higher sensitivity Arecibo observations in Paper II. 

\subsection{The \MHIMstar\ -- \Mstar\ relationship, including \HIit\ non-detections} \label{MHIMstar}
\begin{figure*} 
\centering
\includegraphics[width=18cm]{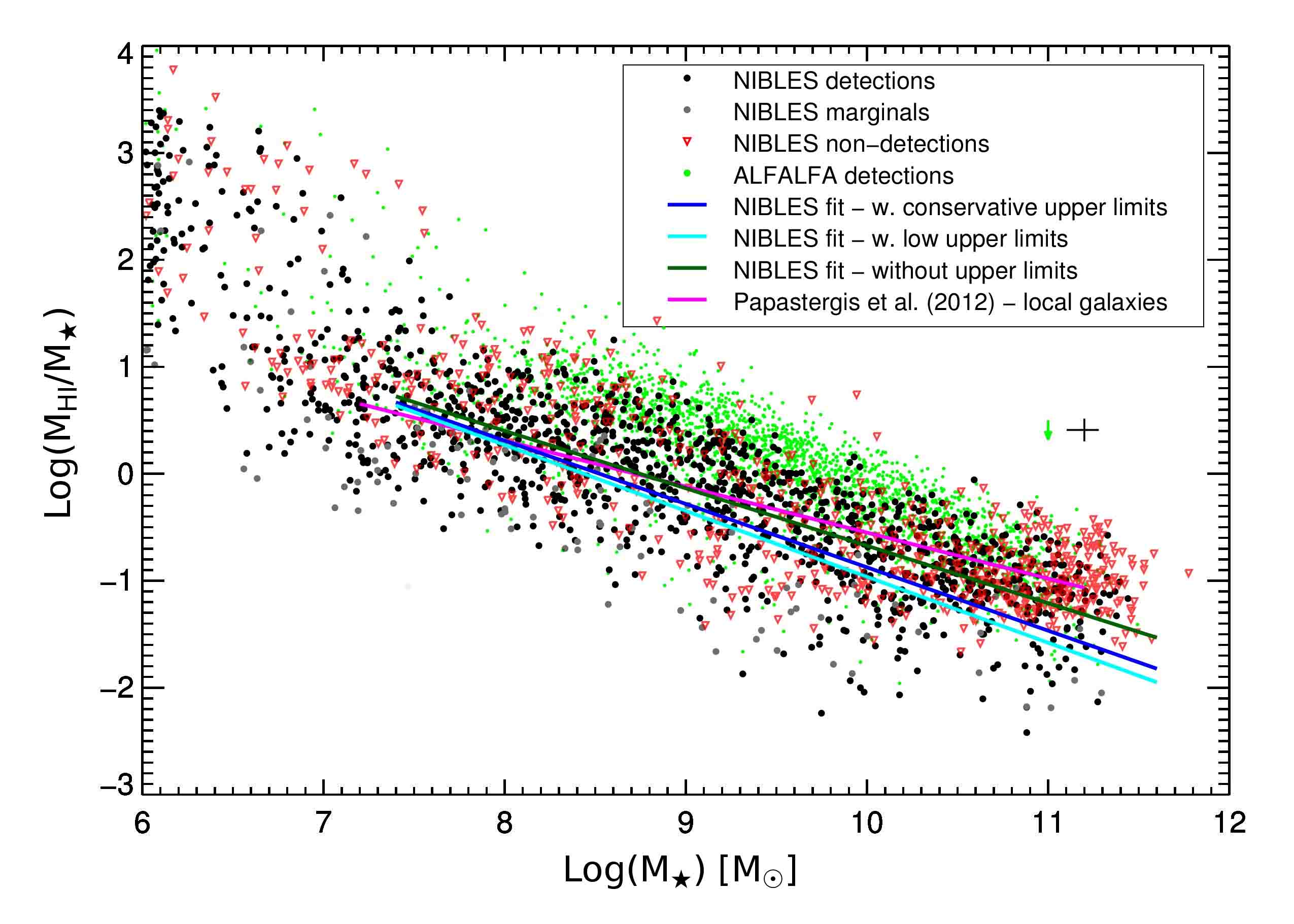} 
\caption{The ratio of total \HI\ and stellar masses, \MHIMstar, as a function of total stellar mass, \Mstar\ (in \Msun). Black dots represent clear NIBLES detections, gray dots marginal detections, and red open triangles estimated upper limits for non-detections, whereas green dots represent the clear ALFALFA detections from \citet{haynes11}. All total stellar masses used for the NIBLES and ALFALFA galaxies were taken from the SDSS ``added-value" MPA/JHU catalogs (see Sect.~\ref{results}). The \HI\ masses of the ALFALFA detections were calculated in the same way as for the NIBLES sources, using simply a distance of $D$ = $V$/H$_0$, where the adopted Hubble constant is $H_0$ = 70 \kms\ Mpc$^{-1}$. The green vertical arrow on the right indicates the mean flux scale difference of 0.16~dex in log(\MHI) between the higher ALFALFA $\alpha$.40 catalog \HI\ line fluxes and the NIBLES values (see Sect.~\ref{comparison}). For all fits made to the NIBLES data, we excluded sources that are either resolved, confused or missing significant SDSS flux. The dark green line shows a linear regression fit made only to the NIBLES detections. The dark blue line shows a linear regression fit to all NIBLES data including our rather conservative estimates for \HI\ mass upper limits based on the upper envelope in the \Wtwenty-\Lr\  relationship (Fig.~\ref{fig:W20Lr}), whereas the light blue line shows a fit using low upper limit estimates based on the mean \Wtwenty-\Lr\  relationship (see text). The pink line shows the mean relationship derived for four literature reference samples of \HI-detected local galaxies from \citet{papastergis12}, which is based on a stellar mass estimation method different from the one used in the MPA/JHU catalogs (see Sect.~\label{MHIMstar}). For the NIBLES detections, we indicate the typical uncertainty in both quantities by the black cross on the right hand side of the plot (see text for details).
}
\label{fig:MHIMstar}
\end{figure*}

The ratio of \HI\ and stellar masses, \MHIMstar, as a function of total stellar mass is shown in Fig.~\ref{fig:MHIMstar} for NIBLES detections, marginals, and estimated upper limits for non-detections. We did not use those sources we flagged as either $C1/C2$ (confused by another galaxy in the NRT beam), $R$ (resolved by the NRT beam) or $U$ (significant SDSS flux missing), as their \MHIMstar\ ratios will be either under- or overestimated by unknown amounts (see Sect.~\ref{results}). 
Overlaid on this plot are high-quality ALFALFA detections, and the mean relationship for four literature reference samples of local \HI-detected galaxies (pink line) from \citet{papastergis12}. We first explore the differences between the plotted properties for the various samples and their uncertainties, both relative and systematic.

Like for the NIBLES data, the total stellar masses used here for the ALFALFA detections were taken by us from the SDSS ``added-value" MPA/JHU catalogs (see Sect.~\ref{results}). To estimate the the uncertainty in the stellar masses of individual galaxies we examined the +1 $\sigma$, -1 $\sigma$ and median mass estimates, and found a typical (mean) relative uncertainty of about 20\%. Matching positions with those of the 15,598 $\alpha$.40 catalog \HI\ detections resulted in 2500 matches. 

A caveat in the comparison between our MPA/JHU catalogs' total stellar masses for NIBLES and ALFALFA galaxies and those used for the reference samples in \citet{papastergis12} is that the latter used a somewhat different way to estimate stellar masses, see \citet{huang12b} for details. We do not have a simple way of estimating the systematic uncertainties between the two stellar mass estimate methods as stellar masses for individual galaxies are not given in \citet{papastergis12}. We will therefore ignore this uncertainty, but we note that in our experience different mass estimates usually agree within about 0.3~dex \citep[e.g.,][]{drory04, moustakas13, sorba15}.

For the NIBLES \HI\ masses, we estimate a typical relative uncertainty, due to variations within the telescope system, of about 15\% and a systematic uncertainty of about 10\%, after comparison with flux scales of other telescopes (see Sect.~\ref{comparison}). Even when correcting the log(\MHI) of the ALFALFA $\alpha$.40 catalog detections by $-$0.16~dex, they lie on, and above, the upper envelope of the NIBLES detections. 

The mean relationship found by Papastergis et al. for the ensemble of four literature samples they use to evaluate the gas-to-stellar mass ratio of galaxies as a function of \Mstar\ (pink line) is log(\MHIMstar) = $-$0.43 log(\Mstar) + 3.75. 

The latter reference samples only contain objects selected based on previously known \HI\ detections, and do not include any \HI\ non-detections. Among the total of about 1000 galaxies, the largest sample used \citep{zhang09} is based on 721 \HI\ detections from HyperLeda, two others are based on Westerbork radio synthesis observations \citep{swaters02, noordermeer05} of 239 galaxies selected to have a strong enough \HI\ line to enable imaging and the smallest \citep{garnett02} contains 42 \HI-detected objects. Their \HI\ masses as used by Papastergis et al. are based on the originally published values, and are independent of subsequent ALFALFA measurements. 

Papastergis et al. also estimated maximum upper limits to \MHIMstar\ ratios for galaxies in their selected sky regions covered by ALFALFA which were not detected by ALFALFA, using \HI\ mass upper limits based on the 25\% completeness limit of the $\alpha$.40 catalog as a function of \HI\ line width. These ratios also lie systematically above the relationship for the four literature samples. 

In order to keep Fig.~\ref{fig:MHIMstar} readable, we did not plot the uncertainties for all individual galaxies but instead indicated the typical uncertainty for both the stellar mass and the \HI-to-stellar mass ratio of the NIBLES detections. For the \HI\ mass, we added the typical relative and systematic uncertainties in quadrature. 

A second, though minor, caveat in comparing samples shown in Fig.~\ref{fig:MHIMstar} concerns the distance scales used. For NIBLES we used a pure Hubble-flow method and heliocentric velocities, whereas for published ALFALFA detections (including those used in \citealt{papastergis12}) a correction for peculiar motions was applied for $V_{\rm CMB}$$<$6000 \kms\ (see \citealt{haynes11} for details). The ALFALFA \HI\ masses shown in Fig.~\ref{fig:MHIMstar} were all calculated by us using the same method as for NIBLES. 

For the four reference samples used in \citet{papastergis12}, no individual distances are given, nor in the paper that comprises the bulk of those data, \citet{zhang09}. From the fact that the latter are all SDSS galaxies with pre-ALFALFA \HI\ detections listed in HyperLeda, we deduce they are relatively nearby objects with a mean velocity of a few thousand \kms. At such velocities, the ALFALFA distances are only about 7\% higher than the NIBLES values, corresponding to +0.03~dex in log(\MHI) -- small compared to the other typical uncertainties of individual galaxy data as shown in Fig.~\ref{fig:MHIMstar}.

A linear regression fit made only to the NIBLES detections (excluding resolved and confused sources) results in a mean relationship (dark green line in Fig.~\ref{fig:MHIMstar}) of log(\MHIMstar) = $-$0.54 log(\Mstar) + 4.70.

We now want to examine the impact of \HI\ non-detections, of which the estimated upper limits to their total \HI\ masses are routinely ignored, in the study of this relationship for our optically-selected NIBLES sample. To this purpose, we fitted our data, including \MHI\ upper limits, with three estimators of the slope and intercept, from the STSDAS statistics package \footnote{http://stsdas.stsci.edu/cgi-bin/gethelp.cgi?statistics}. These were the Buckley-James, expectation-maximization (EM) algorithm, and Schmitt binning methods of linear regression. It is beyond the scope of the paper to discuss the differing nature of these methods \citep[see][for details]{feigelson85, isobe86}. 

At low stellar masses, log(\Mstar)$\approxlt$ 7.5, the relationship between \MHIMstar\ and \Mstar\ apparently becomes non-linear, and we therefore excluded galaxies below this limit from these linear fits. It is sufficient to say here, that all three methods gave similar slopes and intercepts within 0.1~dex in log(\MHIMstar). Using the Buckley-James method, we find log(\MHIMstar) = $-$0.59 log(\Mstar) + 5.05. 

Since we were concerned that our estimated 3$\sigma$ \MHI\ upper limits might be overly conservative owing to our choice of the largest observed \Wtwenty\ line widths as a function of \Lr\  (see Fig.~\ref{fig:W20Lr}), we also explored how adjusting our upper limit estimates would influence the fits. To this end, we subtracted 0.25~dex from the upper envelope line shown in Fig.~\ref{fig:W20Lr}, which corresponds to the mean relationship between \Wtwenty\ and \Lr. In this case, we found log(\MHIMstar) = $-$0.62 log(\Mstar) + 5.20. 

We show these two fits, dark blue when using our conservative upper envelope limits and light blue based on the mean \Wtwenty-\Lr\  relationship, in Fig.~\ref{fig:MHIMstar}. The uncertainties in our fits are 0.03 in the slope and 0.3 in the zero-point. Although \citet{papastergis12} do not provide an estimate of the uncertainties in their fit, the spread of the data points in their Fig. 19 indicates a standard deviation of order $\pm$0.2 dex in log(\MHIMstar) around the mean relationship. 

Overall, our regression fits to all NIBLES data are not very dependent on the choice of 
line widths for \HI\ mass upper limit estimates based on our \nan\ data, 
which on average causes a difference of 0.25 dex in log(\MHI); 
the difference in log(\MHIMstar) between using the conservative, upper envelope widths 
and the mean widths amounts to only 0.03~dex at log(\Mstar) = 7.5 and goes up to 0.12~dex at 11.5. 
About one quarter of the NIBLES galaxies were not detected in \HI, and most of these have the highest stellar masses. Therefore, the effect of using lower \MHI\ estimates for non-detections will have the greatest effect at the high mass end.

On the other hand, the difference between the Papastergis et al. \HI-detected reference samples 
and the NIBLES fit using conservative upper limits amounts to $-$0.08~dex at log(\Mstar) = 7.5 and goes up to 0.57~dex at log(\Mstar) = 11.5. However, keep in mind there may be a systematic difference of up to 0.3~dex between the two methods used for total stellar mass estimates. 

Our follow-up Arecibo detections of 72 \nan\ non-detections show that they lie mainly among the lower envelope of the \nan\ detections and marginals in Fig.~\ref{fig:MHIMstar}, indicating that \HI\ observations sufficiently sensitive to detect all our targets would significantly lower the mean \MHIMstar\ ratio over the entire range of \Mstar\ for our NIBLES sample of optically selected local galaxies. We will discuss this in further detail in future papers in this series.

\begin{acknowledgements} 
We wish to thank the anonymous referee for carefully reading our manuscript and making a number of very useful comments.
We wish to thank Eric G\'erard, and Laurence Alsac and other \nan\ Radio Telescope staff members for their technical advice and assistance. 
TJ and RM are grateful for the bursary provided by the South African SKA Project Office; SB, TJ, RCKK, MR \& PS were supported by the South African National Research Foundation.
The \nan\ Radio Astronomy Station is operated as part of the Paris Observatory, in association with the Centre National de la Recherche Scientifique (CNRS) and partially supported by the R\'egion Centre in France. 
This research has made use of the HyperLeda database (http://leda.univ-lyon1.fr), the NASA/IPAC Extragalactic Database (NED) which is operated by the Jet Propulsion Laboratory, California Institute of Technology, under contract with the National Aeronautics and Space Administration and the Sloan Digital Sky Survey which is managed by the Astrophysical Research Consortium for the Participating Institutions. 
\end{acknowledgements}

\bibstyle{aa}
\bibliographystyle{aa}

\bibliography{NIBLES_bib}



\newpage

\appendix
\section{Notes on individual galaxies}\label{notes}
\begin{itemize}
\item{\bf Source 0023} (\object{KUG 0007+140}): 
Detected in ALFALFA, but not in NIBLES. Small, blueish galaxy of about 
60$^{\circ}$ inclination, without clear spiral arms. The ALFALFA detection has a \Wfifty\ of 205 \kms\ and a 7 mJy peak flux density (peak \SNR\ = 2.8, \SN\ = 6.8, $rms$ = 2.5 mJy). The NRT $rms$ is lower, 1.8 mJy, but the NIBLES spectrum only shows a narrow, 6 mJy peak at the optical velocity. The mean ALFALFA flux density of 5.4 mJy is merely at the 3.0 $\sigma$ level for the NIBLES spectrum. However, our four times deeper Arecibo observations (see Paper II) show a clear detection (\SN\ 15) at a mean level of 3.5 mJy, which is considerably lower than the ALFALFA value, and too low to expect a NIBLES detection.
\item{\bf Source 0117} (\object{SHOC 30}):
Its \Wtwenty\ \HI\ line width of 390 \kms\ is significantly higher
than for the other detections of similar luminosity, log($L_{\rm r}$) =  8.25 (see Fig.~\ref{fig:W20Lr}), but it is a relatively weak detection and its estimated uncertainty in \Wtwenty\ is of order 30 \kms,
which could move it below the upper envelope plotted in Fig.~\ref{fig:W20Lr}.
Also, the detection could be confused, as there are two other blue compact objects in the NRT beam, without known redshifts.
\item{\bf Source 0609} (\object{KUG 0814+251}): 
Its DR9 and DR6 velocities are significantly different (2075 and 1496 \kms, respectively), and we used the mean independent optical velocity of 1888 \kms\ \citep{augarde94}. It has no \HI\ detection, either from NIBLES or in the literature.
\item{\bf Source 0718} (\object{CGCG 263-080}): 
Its DR9 and DR6 velocities are very different (7602 and 2340 \kms, respectively), and we used the independent optical velocity of 7669 \kms\ \citep{falco99}. It has no \HI\ detection, either from NIBLES or in the literature.
\item{\bf Source 0784} (\object{UGC 4722}): 
The \Wfifty\ line width is much smaller in NIBLES than in ALFALFA (57 and 137 \kms, respectively) 
but the ALFALFA spectrum could not be found in the survey's online database. The NRT profile becomes 
much broader just below the 50\% level, and its \Wtwenty\ = 152 \kms.
\item{\bf Source 0793} (\object{NGC 2743}): 
The NIBLES profile is as broad as the ALFALFA detection at the \Wtwenty\ level, but at 
the 50\% level the lower-velocity half of the line is considerable weaker in the NIBLES 
profile, hence the  difference in \Wfifty\ (74 and 178 \kms, respectively). 
The NRT spectrum sloping off could in principle be due to a superimposed, narrow OFF-beam 
detection.
\item{\bf Source 0824} (\object{KUG 0910+433}): 
Its DR9 and DR6 velocities are very different (4232 and 2641 \kms, respectively), and we used the mean independent optical velocity of 4221 \kms\ \citep{saunders00, falco99}. It has no \HI\ detection, either from NIBLES or in the literature.
\item{\bf Source 0828} (\object{ASK 261057}): 
Its DR9 radial velocity is unconstrained (nominally 1245 \kms), 
and we adopted our Arecibo NIBLES \HI\ value of 1370 \kms\ (Paper II). There are no independent literature values.
\item{\bf Source 0958} (\object{PGC 4571034}): 
The \Wfifty\ line width is much larger in NIBLES than in ALFALFA 
(116 and 32 \kms, respectively). The peak in the NIBLES profile which is centered on the 
optical velocity has about the same width as the ALFALFA profile, but in addition it has 
a noise peak centered on a $\sim$80 \kms\ higher velocity which rises up to 60\% of the 
main peak level. The latter can explain the measured width difference. 
We could not identify a potentially confusing other galaxy within the NRT beam.
\item{\bf Source 0998} (\object{NGC 3156}): 
Its SDSS DR9 radial velocity is unconstrained (nominally 1245 \kms), and 
we adopted the mean independent optical value of 1266 \kms\ \citep{sandage78,tonry81,vennik82,falco99}.
It has no \HI\ detection, either from NIBLES or in the literature.
\item{\bf Source 1319} (\object{PGC 1275866}): 
The \Wfifty\ line width is much larger in NIBLES than in ALFALFA 
(69 and 37 \kms, respectively). The peak in the NIBLES profile which is centered on the 
optical velocity has about the same width as the ALFALFA profile, but in addition it has 
a noise peak centered on a $\sim$100 \kms\ lower velocity which rises up to 35\% of the 
main peak level. The latter can explain the measured width difference. We could not 
identify a potentially confusing other galaxy within the NRT beam.
\item{\bf Source 1355} (\object{MCG +09-19-160}): 
Its SDSS DR9 redshift is unconstrained (nominally at 1245 \kms), and we 
adopted the value of 1233 \kms\ of the marginal NIBLES \HI\ detection, which is marginally lower than the independent optical velocity of 1322$\pm$50 \kms\ \citep{falco99}.
\item{\bf Source 1371} (\object{CGCG 292-024}): 
Its SDSS DR9 radial velocity is unconstrained (nominally 1245 \kms), and 
we adopted the NIBLES \HI\ value of 1294 \kms\ which agrees with the independent optical literature value of 1265 \kms\ \citep{sandage78, tonry81, vennik82, falco99} and the previous \nan\ detection at 1282 \kms\ \citep{garcia94}.
\item{\bf Source 1598} (\object{VCC 0315}): 
Its DR9 and DR6 velocities are significantly different (1589 and 1461 \kms, respectively), and we used the independent optical velocity of 1575 \kms\ \citep{grogin98, falco99}. It has no \HI\ detection, either from NIBLES or in the literature.
\item{\bf Source 1654} (\object{PGC 40315}): 
Its SDSS DR9 radial velocity is unconstrained (nominally 1245 \kms), and 
we adopted its mean independent optical value of 1370 \kms\ (Prugniel 2001, private comm.; see HyperLeda). It has no \HI\ detection, either from NIBLES nor in the literature.
\item{\bf Source 1715} (\object{IC 3665}): 
Its DR9 and DR6 velocities are very different (7 and 1200 \kms, respectively), 
and we used the independent optical velocity of 1227 \kms\ \citep{binggeli85}. It has no \HI\ detection, either from NIBLES or in the literature.
\item{\bf Source 1717} (\object{NGC 4649}): 
Its DR9 and DR6 velocities are different (1169 and 1018 \kms, respectively) and based on two spectroscopic targets on either side of the galaxy center. Instead, we used the mean independent optical velocity of 1117 \kms\ (from a dozen references, see HyperLeda for details). It has no \HI\ detection, either from NIBLES or in the literature.
\item{\bf Source 1723} (\object{ASK 77777}): 
Its SDSS DR9 radial velocity is unconstrained (nominally 1245 \kms), and 
we adopted the independent optical value of 1169 \kms\ \citep{colless03}. It has no \HI\ detection, either from NIBLES or in the literature.
\item{\bf Source 1771} (\object{PGC 1132599}): 
Its DR9 and DR6 velocities are significantly different (1235 and 982 \kms, respectively), and we used the independent optical velocity of 1229 \kms\ \citep{colless03}. It has no \HI\ detection, either from NIBLES or in the literature.
\item{\bf Source 1970} (\object{NGC 5356}): 
The NIBLES and ALFALFA profiles are as broad at the \Wtwenty\ level, 
but the NIBLES profile is much narrower at the \Wfifty\ level (127 vs. 278 \kms, respectively), 
which appears to be due to a narrow OFF-beam detection centered on $\sim$1320 \kms\ 
and cutting through the lower-velocity half of the line profile.
\item{\bf Source 2140} (\object{ASK 082514}): 
Its DR9 radial velocity is unconstrained (nominally 1245 \kms), 
and we adopted our Arecibo NIBLES \HI\ value of 1728 \kms\ (Paper II). There are no independent literature values.
\item{\bf Source 2155} (\object{NGC 5644}): 
Its DR9 and DR6 velocities are very different (7650 and 64 \kms, respectively), and we used the mean independent optical velocity of 7649 \kms\ \citep{huchra83, falco99}. It has no \HI\ detection, either from NIBLES or in the literature.
\item{\bf Source 2167} (\object{NGC 5675}): 
Its \Wtwenty\ \HI\ line width of $\sim$1030 \kms\ is significantly higher than for the other detections of similar luminosity, log(\Lr) = 10.4 (see Fig.~\ref{fig:W20Lr}), but the SDSS image clearly shows it is an ongoing major merger, so we can expect unsettled gas at extreme velocities.
\item{\bf Source 2183} (\object{LSBC D723-05}): 
Although the NIBLES and ALFALFA \HI\ profiles look similar, the fitted 
\Wfifty\ width is much smaller in NIBLES compared to ALFALFA 
(62 and 121 \kms, respectively). This may be due to the bumpy slope of the 
low-velocity edge of the profile, the presence of noise spikes and differences 
in the line width measurement algorithms.
\item{\bf Source 2265} (\object{PGC 3350778}): 
Its SDSS DR9 radial velocity is unconstrained (nominally 1245 \kms), and 
we adopted the independent optical value of 1312 \kms\ \citep{mahdavi05}. It has no \HI\ detection, either from NIBLES or in the literature.
\item{\bf Source 2356} (\object{CGCG 137-019}): 
Its DR9 and DR6 velocities are very different (4508 and 12 \kms, respectively), and we used the mean independent optical velocity of 4555 \kms\ \citep{falco99}. It has no \HI\ detection, either from NIBLES or in the literature.
\item{\bf Source 2555} (\object{KUG 2335+148}): 
Detected in ALFALFA, but only marginally in NIBLES. Very small blue, roundish galaxy with an arc of star formation clumps; it looks like a late-stage merger. The ALFALFA detection has a \Wfifty\ of 106 \kms\ and peak flux density of 10 mJy (peak \SNR\ = 5, \SN\ = 8.0, $rms$ = 2.1 mJy). The NIBLES $rms$ noise level is lower, 1.7 mJy, but nothing is visible in the NRT spectrum although the mean Arecibo flux density is at the 4.3$\sigma$ level for the NIBLES spectrum.

\end{itemize}

\onecolumn


\newpage
\clearpage

\input{NIBLES_paperone_table1_detections_160710.tex}

\newpage
\clearpage

\input{NIBLES_paperone_table2_marginals_160710.tex}

\newpage
\clearpage

\input{NIBLES_paperone_table3_nondetections_160710.tex}

\newpage
\clearpage

 \input{NIBLES_paperone_table4_fluxcomp_160607.tex}

\newpage
\clearpage

\input{NIBLES_paperone_table5_lowvels_160617.tex}

\newpage
\clearpage

\input{NIBLES_paperone_table6_excluded_160428.tex}

\newpage
\clearpage

\input{NIBLES_paperone_table7_HIfluxcalibrators_250416.tex}

\end{document}

%% file: NIBLES_paperone_table2_marginals_160710.tex
{\tiny
\begin{landscape}
\begin{longtable}{lrrlrrrrrrrrrrrrrr}
\caption{Basic optical and \HI\ data -- marginally detected galaxies  \label{tab:marginals}} \\
\hline\hline
Source & RA & Dec                      & Other name & \Vopt\ & $g$-$z$  & $M_{\rm g}$ & \Mstar\ & \sSFR\ & $rms$ & \VHI\ & \Wfifty\ & \Wtwenty\ & \FHI\  & \SNR\ & \SN\ & \MHI\  & \fMHIMstar\ \\  
       & \multicolumn{2}{c}{(J2000.0)} &            &           &      &       & $[$log$]$  & $[$log$]$ &     &          &          &          &           &     &     & $[$log$]$ & $[$log$]$   \\
       &    &                       &            &  km/s     &  mag & mag   & \Msun\     & yr$^{-1}$ & mJy & km/s     & km/s     & km/s     & Jy km/s   &     &     & \Msun\    &          \\
\hline
\hline
\endfirsthead
\caption{-- {\it continued}.}\\
\hline\hline
Source & RA & Dec                      & Other name & \Vopt\ & $g$-$z$  & $M_{\rm g}$ & \Mstar\ & \sSFR\ & $rms$ & \VHI\ & \Wfifty\ & \Wtwenty\ & \FHI\  & \SNR\ & \SN\ & \MHI\  & \fMHIMstar\ \\  
       & \multicolumn{2}{c}{(J2000.0)} &            &           &      &       & $[$log$]$  & $[$log$]$ &     &          &          &          &           &     &     & $[$log$]$ & $[$log$]$   \\
       &    &                       &            &  km/s     &  mag & mag   & \Msun\     & yr$^{-1}$ & mJy & km/s     & km/s     & km/s     & Jy km/s   &     &     & \Msun\    &          \\
\hline
\hline
\endhead
\hline
\endfoot
      0024      & 00 10 52.65 &  15 35 26.60  & 2MASX J00105266+1535261 & 11071$\pm$6 &  2.19 & -20.07 & 11.29 &   -11.27 & 2.75 & 11053$\pm$10 &  38 &  61 &  0.28$\pm$0.19 &   3.0 &    2.8 &  9.25 & -2.05 \\ 
      0032      & 00 13 43.98 &  00 22 18.20  &               ASK 28319 &  3968$\pm$2 &  0.36 & -15.97 &  7.81 &    -9.28 & 3.01 &  3968$\pm$53 &  28 & 156 &  0.49$\pm$0.34 &   3.5 &    5.1 &  8.58 &  0.78 \\ 
      0036      & 00 15 00.08 & -11 08 03.90  &              ASK 125926 &  3351$\pm$2 & -0.09 & -15.63 &  8.05 &    -9.78 & 2.07 &  3350$\pm$6 & 138 & 149 &  0.29$\pm$0.23 &   2.4 &    2.0 &  8.20 &  0.14 \\ 
    0037$^{A1}$ & 00 15 05.40 &  15 03 59.70  & 2MASX J00150533+1503597 &  5428$\pm$2 &  1.37 & -19.15 & 10.00 &   -10.55 & 1.44 &  5406$\pm$35 & 130 & 192 &  0.21$\pm$0.18 &   2.5 &    2.2 &  8.48 & -1.52 \\ 
      0108      & 00 41 42.50 &  14 17 56.20  &               ASK 40372 &  8810$\pm$2 &  0.61 & -18.15 &  6.66 &    -8.59 & 1.59 &  8847$\pm$13 & 103 & 129 &  0.21$\pm$0.16 &   2.7 &    2.2 &  8.93 &  2.27 \\ 
    0116$^{C1}$ & 00 46 05.70 & -09 30 10.80  &                 IC 0050 &  6027$\pm$2 &  1.49 & -20.54 & 10.79 &   -12.56 & 1.28 &  5969$\pm$12 & 371 & 404 &  0.65$\pm$0.23 &   3.7 &    4.3 &  9.04 & -1.75 \\ 
      0126      & 00 51 49.40 &  00 33 53.10  &                  UM 283 &  4629$\pm$1 &  0.41 & -17.77 &  8.53 &    -9.22 & 1.51 &  4636$\pm$12 & 103 & 121 &  0.22$\pm$0.15 &   2.1 &    2.4 &  8.38 & -0.15 \\ 
      0129      & 00 53 19.60 & -10 24 11.80  & 2MASX J00531961-1024115 &  4403$\pm$1 &  0.46 & -18.06 &  8.80 &    -9.26 & 1.42 &  4404$\pm$29 &  73 & 136 &  0.31$\pm$0.15 &   3.2 &    4.1 &  8.46 & -0.35 \\
\end{longtable}
\end{landscape}
}

%% file: NIBLES_paperone_table3_nondetections_160710.tex
{\tiny
\begin{longtable}{lrrlrrrrrrrrr}
\caption{Basic optical and \HI\ data -- non-detected galaxies  \label{tab:nondetections}} \\
\hline\hline
Source & RA & Dec                      & Name & \Vopt\ & $g$-$z$ & $M_{\rm g}$ & \Mstar\   & \sSFR\      & $rms$ & \SNR\ & \MHI\     & \fMHIMstar\ \\  
       & \multicolumn{2}{c}{(J2000.0)} &      &        &         &             & $[$log$]$ & $[$log$]$ &     &     & $[$log$]$ & $[$log$]$   \\
       &    &                          &      &  km/s  & mag     & mag         & \Msun\    & yr$^{-1}$ & mJy &     & \Msun\    &             \\
\hline
\endfirsthead
\caption{-- {\it continued}.} \\
\hline\hline
Source & RA & Dec                      & Name & \Vopt\ & $g$-$z$ & $M_{\rm g}$ & \Mstar\   & \sSFR\      & $rms$ & \SNR\ & \MHI\     & \fMHIMstar\ \\  
       & \multicolumn{2}{c}{(J2000.0)} &      &        &         &             & $[$log$]$ & $[$log$]$ &     &     & $[$log$]$ & $[$log$]$   \\
       &    &                          &      &  km/s  & mag     & mag         & \Msun\    & yr$^{-1}$ & mJy &     & \Msun\    &             \\
\hline
\endhead
\hline
\endfoot
      0003      & 00 00 47.90 &  14 16 39.10  &    2MASX J00004789+1416390 & 10847$\pm$4 &  1.66 & -20.01 & 10.71 &   -10.49 & 2.76 &   1.7 & $<$10.40 & $<$-0.31 \\ 
      0009      & 00 03 22.20 & -10 46 14.20  &             MCG -02-01-012 &  8937$\pm$2 &  1.57 & -20.87 & 10.98 &   -12.37 & 4.85 &   1.5 & $<$10.55 & $<$-0.43 \\ 
      0012      & 00 04 35.21 &  00 50 54.90  &    2MASX J00043518+0050549 &  6392$\pm$5 &  1.38 & -17.81 &  9.57 &   -10.61 & 1.46 &   2.0 &  $<$9.49 & $<$-0.09 \\ 
      0014      & 00 06 19.61 &  14 19 38.70  &    2MASX J00061957+1419389 &  5448$\pm$3 &  1.31 & -18.22 &  9.66 &   -11.16 & 2.22 &   3.9 &  $<$9.56 & $<$-0.10 \\ 
      0015      & 00 06 29.29 &  14 10 56.40  &    2MASX J00062933+1410559 &  5324$\pm$6 &  1.59 & -17.84 &  9.74 &   -11.30 & 2.44 &   1.8 &  $<$9.56 & $<$-0.18 \\ 
      0018      & 00 08 00.98 &  14 01 18.60  &                 ASK 146993 &  5429$\pm$4 &  0.76 & -17.38 &  8.74 &    -9.73 & 1.95 &   2.9 &  $<$9.41 &  $<$0.67 \\ 
    0023$^{A1}$ & 00 10 25.50 &  14 17 23.00  &               KUG 0007+140 &  5452$\pm$4 &  0.93 & -18.81 &  9.50 &   -10.17 & 1.79 &   0.9 &  $<$9.50 & $<$-0.00 \\ 
      0025      & 00 11 43.20 &  14 28 01.10  &                 ASK 146943 &  5213$\pm$2 &  0.66 & -17.47 &  8.67 &    -9.48 & 2.44 &   1.7 &  $<$9.48 &  $<$0.81 \\
\end{longtable}
}

%% file: NIBLES_paperone_table4_fluxcomp_160607.tex
\begin{table}
\caption{\HI\ surveys flux scale comparison}  
\label{tab:fluxcomp}
\begin{tabular}{lllr}  
\hline\hline         
survey 1 & survey 2 & mean \FHI\ ratio & $N$ \\
\hline
\multicolumn{4}{l}{} \\ 
\multicolumn{4}{l}{single-feed / single-feed detectors} \\
\multicolumn{4}{l}{} \\  
Arecibo                     & \nan\ calibrators           & 1.13$\pm$0.12 &  12 \\
GBT                         & \nan\ calibrators           & 0.93$\pm$0.24 &   2  \\
Arecibo                     & GBT                         & 0.99$\pm$0.18 &  16  \\
\multicolumn{4}{l}{} \\
\end{tabular}   
\end{table}      
                                                                                                 

%% file: NIBLES_paperone_table5_lowvels_160617.tex
{\tiny
\begin{landscape}
\begin{table}
\caption{Basic optical and \HI\ data -- galaxies with V$<$850 \kms\ not used for further analysis} \label{tab:lowvels} 
\begin{tabular}{rrrrlrrrrrrrrrrrrr}
\hline\hline
SOU & RA          & Dec             & Name       & \Vopt\ & $g$-$r$ & $M_g$ & \Mstar\   & $sSFR$    & $rms$ & \VHI\ & \Wfifty\ & \Wtwenty\ & \FHI\   & $SNR$ & $S/N$ & \MHI\  & \fMHIMstar\  \\  
    & \multicolumn{2}{c}{(J2000.0)} &            & DR9    &         &       & $[$log$]$ & $[$log$]$ &       &       &          &           &         &     &         & $[$log$]$ & $[$log$]$    \\
    &                               &            & km/s   & mag     &  mag  & \Msun\    & yr$^{-1}$ & mJy   & km/s  & km/s     & km/s      & Jy km/s &     &         & \Msun\    &              \\
\hline
0347$^{N}$  & 03 06 46.80 &  00 28 10.30  & \object{PGC 1166738} & 711$\pm$7  & 0.27 & -12.31 & 6.82 &    --- & 3.03 & 699$\pm$13 &  38 &  79 &  0.55$\pm$0.25 &  4.6 &   4.9 & 7.11 & 0.30 \\ 
1572$^{A1}$ & 12 13 48.30 &  12 41 25.90  & \object{IC 3052}     & 812$\pm$14 & 0.38 & -13.60 & 7.03 & -10.16 & 3.59 & 834$\pm$9  &  85 & 104 &  0.65$\pm$0.34 &  3.2 &   3.3 & 7.34 & 0.31 \\ 
1734$^{N}$  & 12 49 11.40 &  03 23 15.70  & \object{NGC 4701}    & 755$\pm$3  & 0.88 & -17.27 &  --- &    --- & 2.69 & 719$\pm$1  & 163 & 187 & 30.28$\pm$0.34 & 72.5 & 146.3 & 8.88 &  --- \\ 
1897$^{A1}$ & 13 39 22.34 &  31 14 57.60  & \object{ASK 526959}  & 703$\pm$2  & 0.28 & -13.50 & 6.72 &  -9.14 & 2.45 & 699$\pm$3  &  51 &  81 &  1.79$\pm$0.20 & 13.1 &  16.9 & 7.62 & 0.90 \\ 
2259        & 14 59 01.99 &  44 10 33.40  & \object{ASK 402895}  & 639$\pm$5  & 0.38 & -12.43 &  --- &    --- & 2.30 & 632$\pm$11 &  38 &  66 &  0.32$\pm$0.17 &  3.7 &   3.7 & 6.79 &  --- \\
\hline
\end{tabular}
\end{table}
\end{landscape}                              
}

%% file: NIBLES_paperone_table6_excluded_160428.tex
\begin{table}
\caption{Basic optical and \HI\ data -- selected galaxies with redshift problems not usable for further analysis} 
\label{tab:excluded}
\begin{tabular}{rrlll}
\hline\hline
RA          & Dec             & Name        & \Vopt\     & \Vopt\       \\  
\multicolumn{2}{c}{(J2000.0)} &             & DR9        & other        \\
            &                 &             & (km/s)     & (km/s)       \\
\hline
\multicolumn{5}{l}{} \\
\multicolumn{5}{l}{DR9 redshift unconstrained:} \\
\multicolumn{5}{l}{} \\
02 21 14.10 & -00 53 12.1  &   \object{ASK 35835}    & 1245?     & [DR6: 1309]   \\ 
02 33 31.90 & -00 44 38.0  &   \object{ASK 36025}    & 1245?     & [DR6: 1541]   \\
\hline
\end{tabular}
\end{table}

%% file: NIBLES_paperone_table7_HIfluxcalibrators_250416.tex
\begin{table}
\caption{\HI\ line flux calibrator galaxies} 
\label{tab:fluxcalibgals}
\begin{tabular}{lrrr}  
\hline\hline         
Name       &  Arecibo    & GBT         &  NRT         \\ 
           & \FHI\       & \FHI\       & \FHI\        \\
           & (Jy km/s)   & (Jy km/s)   & (Jy km/s)    \\
\hline
\multicolumn{4}{l}{} \\
\object{UGC  1246} &  8.6$\pm$0.1 &             &  7.1$\pm$0.4 \\   
\object{UGC  2432} &  5.9$\pm$0.1 &             &  5.1$\pm$0.3 \\   
\object{UGC  2809} &  9.9$\pm$0.1 &             &  9.8$\pm$0.3 \\   
\object{UGC  3755} &  8.2$\pm$0.1 &             &  7.8$\pm$0.3 \\   
\object{UGC  3946} & 12.0$\pm$0.1 &             & 10.9$\pm$0.7 \\   
\object{UGC  4117} &  4.1$\pm$0.1 &             &  3.2$\pm$0.1 \\   
\object{UGC  4660} &  6.0$\pm$0.1 &             &  5.4$\pm$0.1 \\   
\object{UGC  5160} &  2.9$\pm$0.2 & 2.9$\pm$0.3 &              \\ 
\object{UGC  5215} &  6.8$\pm$0.5 & 7.5$\pm$0.8 &              \\      
\object{UGC  5218} &  6.7$\pm$0.1 & 7.7$\pm$0.8 &              \\      
\object{UGC  6886} &  4.6$\pm$0.2 & 4.8$\pm$0.5 &              \\      
\object{UGC  7976} &  6.4$\pm$0.1 &             &  5.9$\pm$0.3 \\   
\object{UGC  8091} &  7.0$\pm$0.1 & 8.8$\pm$0.9 &              \\      
\object{UGC  8249} &  7.5$\pm$0.1 & 9.0$\pm$0.9 &              \\     
\object{UGC  8503} &  3.7$\pm$0.1 &             &  3.0$\pm$0.1 \\   
\object{UGC  9007} &  1.4$\pm$0.1 & 1.5$\pm$0.2 &              \\      
\object{UGC 11578} &  8.7$\pm$0.1 &             &  9.2$\pm$0.9 \\   
\object{UGC 11926} &  5.2$\pm$0.1 &             &  4.2$\pm$0.3 \\   
\object{UGC 11992} &  5.3$\pm$0.1 & 6.2$\pm$0.6 &              \\      
\object{UGC 12695} &  4.2$\pm$0.1 &             &  3.6$\pm$0.2 \\   
\multicolumn{4}{l}{} \\
\hline
\multicolumn{4}{l}{{\bf Notes}: Arecibo (at 8.5 \kms\ resolution) and} \\
\multicolumn{4}{l}{GBT values are from \citet{oneil04HI} } \\
\end{tabular}   
\end{table}      
            